\documentclass[conference]{IEEEtran}
\IEEEoverridecommandlockouts
\usepackage{cite}
\usepackage{amsmath,amssymb,amsfonts}
\usepackage{graphicx}
\usepackage{textcomp}
\usepackage{xcolor}
\usepackage{url}
\usepackage{cite}
\usepackage{algorithm}
\usepackage{algpseudocode}

\definecolor{hosseinColor}{HTML}{1F77B4}

\def\BibTeX{{\rm B\kern-.05em{\sc i\kern-.025em b}\kern-.08em
    T\kern-.1667em\lower.7ex\hbox{E}\kern-.125emX}}
\begin{document}

\title{The Promise of Strategic Agentic Scambaiting}

\author{
\IEEEauthorblockN{
Abhishek Satyam\IEEEauthorrefmark{1},
Hossein Siadati\IEEEauthorrefmark{2}\IEEEauthorrefmark{3},
Haadi Jafarian\IEEEauthorrefmark{3},
Sima Jafarikhah\IEEEauthorrefmark{4},
Juntao Chen\IEEEauthorrefmark{1}
}
\IEEEauthorblockA{
\IEEEauthorrefmark{1}Fordham University, USA\\
\IEEEauthorrefmark{2}Cybera Global, USA\\
\IEEEauthorrefmark{3}University of Colorado Denver, USA\\
\IEEEauthorrefmark{4}University of North Carolina Wilmington, USA
}
\IEEEauthorblockA{
as526@fordham.edu,
s.h.siadaty@gmail.com,
haadi.jafarian@ucdenver.edu,\\
jafarikhaht@uncw.edu,
jchen504@fordham.edu
}
}

\maketitle



\begin{abstract}

Scammers have increasingly adopted AI-driven capabilities, evolving from automated message generation to fully agentic conversational systems that scale operations and maximize the exploitation of fraudulent resources before detection and takedown. In parallel, agentic scambaiting has emerged as a defensive approach in which autonomous agents impersonate potential victims and engage scammers in sustained interactions to extract actionable threat intelligence, such as mule bank accounts and cryptocurrency wallet addresses, that can be reported to disrupt fraudulent infrastructures. Together, these developments transform online fraud into an adversarial socio-technical ecosystem in which autonomous scamming and scambaiting agents engage in deceptive interactions while competing over the discovery, preservation, and exploitation of one another’s resources. A fundamental and largely unexplored question is how defensive scambaiting can remain effective in an agent-to-agent environment, and under what conditions either side gains a strategic advantage. In this paper, we present a formal model of agent-to-agent fraud interactions that captures the strategic decision spaces available to both attackers and defenders. Using this framework, we analyze the dynamics of engagement, resource consumption, belief adaptation, and disruption outcomes.  We find that scammer profitability turns negative once baiting-agent prevalence exceeds certain thresholds, and that when both sides improve at the same pace, defender investment in mimicry quality outpaces equivalent gains in scammer detection capability, so the arms race favors the defender under our baseline parameters. Our model and findings provide insights into the stability, scalability, and long-term effectiveness of autonomous defensive agents in an evolving AI-mediated fraud landscape.

\end{abstract}

\section{Introduction}

Online fraud losses continue to increase rapidly, reflecting a sustained shift in the way scams are executed and monetized. According to Federal Trade Commission (FTC) data, U.S.\ consumers reported \$15.9\,billion in losses in 2025, a substantial increase from \$12\,billion the prior year~\cite{FTC2026FraudLosses}. When accounting for underreporting, the FTC estimated the overall cost of fraud to consumers for 2024 could be as high as\$195.9\,billion~\cite{FTC2026FraudLosses}. The FBI's Internet Crime Complaint Center (IC3) similarly recorded \$20.9\,billion in cyber-enabled losses in 2025, the first year complaints surpassed one million, representing a 26\% year-over-year increase~\cite{fbiannualinternetcrimereport2025}. Much of this growth is driven by Authorized Push Payment (APP) scams, where victims are manipulated into initiating payments themselves, typically via bank transfers or cryptocurrency~\cite{FTC2026FraudLosses,experianfutureoffraudforecast2026}. APP scam losses across six major real-time payment markets are projected to reach \$7.6\,billion by 2028, at a compound annual growth rate of 12\%~\cite{ACIWorldwide_Scamscope}. Unlike traditional account takeover fraud, APP scams are difficult to detect because payments are authorized by victims, making conventional indicators (e.g., credentials, devices, cookies, and timing) appear legitimate. Stolen funds are commonly routed through mule accounts belonging to unwitting individuals recruited via job offers, romance scams, or other social-engineering schemes~\cite{florencio2010mules}. Recovery is further constrained by fast payment systems and rapid laundering: a 2024 U.S.\ Senate investigation found that the three largest Zelle owner banks reimbursed victims of APP scams only 12\% of the time in 2023, while victims of unauthorized account fraud were reimbursed only 38\% of the time, down from 62\% in 2019~\cite{SenatePSI2024Zelle}. These challenges highlight the need to move 
beyond user education and incremental fraud detection toward disrupting the  operational infrastructure of scam ecosystems.
 
\begin{figure}[t]
    \centering
    \includegraphics[width=\linewidth]{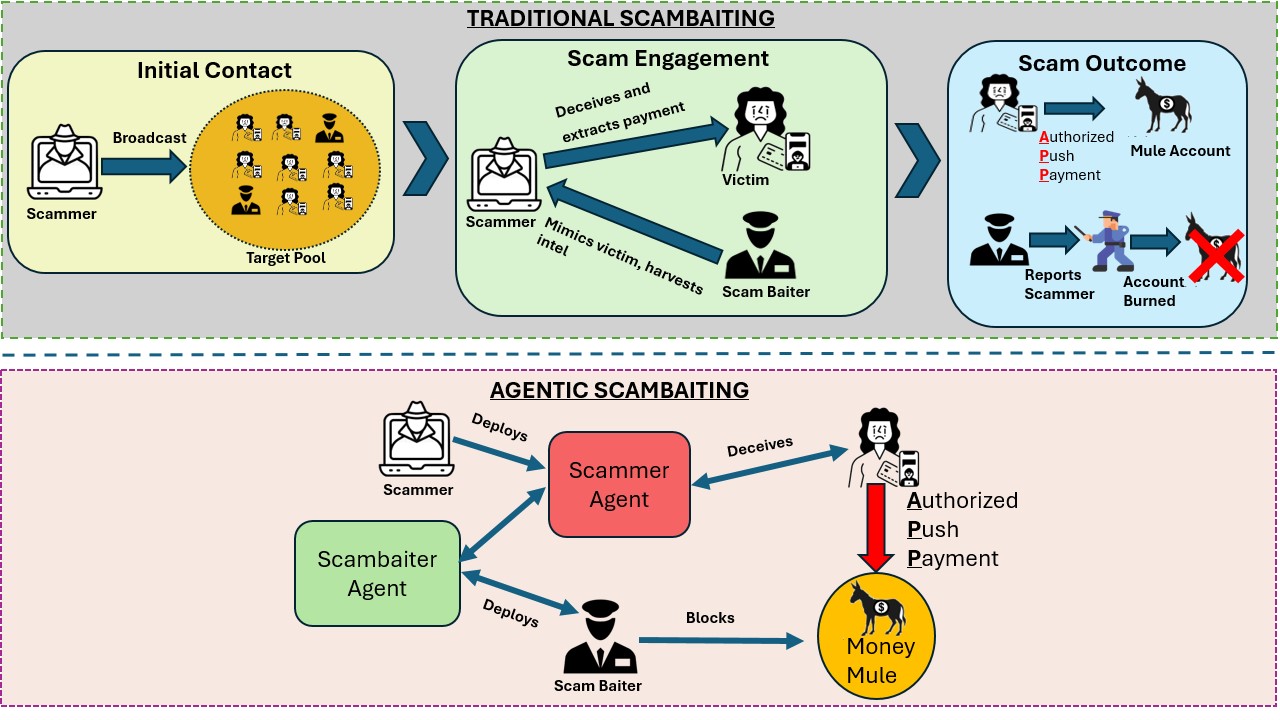}
    \caption{Traditional vs.\ agentic scambaiting 
    }
    \label{fig:traditional-vs-agentic}
\end{figure}

Building on this need for actionable CTI, an emerging approach uses agentic conversational engagement, where LLM agents simulate realistic victim or accomplice personas to interact with scammers~\cite{Netsafe2017, cybera2026, beesafeLabs2026, hackaday2025aiScammers, apateAI2025, foxnews2024AIPoweredGrandma, canham2022planting}. This extends traditional scambaiting, in which security enthusiasts manually engaged with scammers. Meanwhile, scammers increasingly leverage generative AI to automate and enhance scam conversations~\cite{roy2024phishbots, afane2024next}, as well as deepfakes for convincing real-time audio and video impersonation attacks~\cite{wenger2021hello}. Figure~\ref{fig:traditional-vs-agentic} illustrates this evolution of scambaiting.
The growing use of agentic models by both scammers and defenders turns scam engagement into a strategic interaction rather than a one-sided automation problem. Scammers can use conversational cues to screen for non-viable targets and avoid exposing fragile infrastructure, while defenders must sustain credible engagement long enough to induce disclosure. A key structural insight motivating our model is that scam conversations progress through a staged kill chain, moving from trust-building through escalation to a request phase and finally to disclosure, and the mule account details that constitute actionable threat intelligence only surface at the late disclosure stage. Stage advancement is not automatic but is driven by the scammer's evolving belief that the counterparty is a genuine victim worth the escalating commitment. This means scambaiting is fundamentally a screening survival problem in which a baiting agent must avoid triggering suspicion through the early stages before the intelligence it seeks becomes available. This motivates a formal treatment of scambaiting not merely as an engineering system, but as an adaptive security interaction in which both sides respond to one another's perceived capabilities and behavior.

The central challenge we study is how the parallel adoption of AI by both sides transforms scambaiting into an adversarial arms race in which scammers deploy increasingly capable agentic systems to automate screening and distinguish baiters from real victims, while defenders must continuously improve their agents to survive that screening and extract actionable intelligence. By modeling these strategies formally, we examine when scambaiting continues to generate meaningful intelligence, when attacker adaptation reduces its effectiveness, and which structural conditions determine the long-run outcome. We answer these questions through the following contributions:

\begin{itemize}
    \item \textbf{A formal model of agentic scambaiting.} We formalize 
    each engagement as a finite-horizon Dynamic Bayesian Game in which 
    the scammer's decisions are driven by an evolving belief about whether 
    the counterparty is a genuine victim or a defender-controlled baiting 
    agent, capturing the asymmetry that makes screening both rational and 
    costly. The model's staged structure, signal design, and payoff 
    ordering are informed by operational patterns documented in a 
    large-scale LLM-based scambaiting deployment~\cite{siadati2025send}, 
    grounding the formal framework in observed engagement dynamics.

    \item \textbf{A belief-driven framework for conversational screening and adaptation.}
    We model scammer-side inference as a sequential belief-updating process in which attackers distinguish scambaiters from genuine victims through noisy conversational cues. The framework captures how scammer detection capability and defender mimicry jointly shape engagement duration, abandonment behavior, and the likelihood of actionable intelligence disclosure.

    \item \textbf{A campaign-level resource attrition framework.} We extend the single-episode model to repeated interactions in which scammer mule accounts and defender email identities are finite pools that erode over time, enabling analysis of long-run sustainability that per-episode models cannot capture.

    \item \textbf{Computational characterization of deployment conditions.} Through backward induction and Monte Carlo simulation, we identify the deployment scales at which scamming becomes economically unviable, the conditions under which attacker screening causes intelligence yield to collapse, and the design properties that sustain defensive value as both sides adapt.
\end{itemize}

The remainder of this paper is organized as follows. Section~\ref{sec:related} reviews prior work on scambaiting, AI-driven adversarial engagement, and strategic adaptation in cybersecurity. Section~\ref{sec:problem} presents the problem statement and research questions motivating this study. Section~\ref{sec:foundation} introduces the foundational framework for strategic scambaiting, including the engagement model, belief update process, and resource attrition dynamics. Section~\ref{sec:formal_model} describes the computational approach used to compute equilibrium strategies and simulate large-scale interactions. Section~\ref{sec:simulation} presents the experimental setup and empirical grounding of the model. Section~\ref{sec:results} reports the evaluation results and analyzes the conditions under which agentic scambaiting remains effective or degrades under attacker adaptation. Finally, Section~\ref{sec_conclusion} discusses implications, limitations, and future research directions.

\section{Related Work}
\label{sec:related}
Historically, cyber adversarial behavior has evolved in response to incentives, operational constraints, and deployed defenses. This is seen in spam ecosystems shaped by economic return and campaign efficiency \cite{kanich2008spamalytics} and in CAPTCHA-solving markets undermined by adaptive cost-optimizing services \cite{motoyama2010recaptchas}. Scambaiting is not an exception. Scammers are not passive targets of automation. They screen interactions, adapt their behavior, and disengage when conversations appear suspicious or unprofitable. To better understand this evolving landscape, we examine prior literature through traditional scambaiting, AI-driven scambaiting, and strategic models of adversarial adaptation.

\subsection{Scambaiting as Adversary Engagement}
\label{subsec:trad_scambaiting}

Scambaiting predates modern language models. Instead of blocking scam attempts at first contact, defenders sustain interactions while posing as plausible victims to consume attacker resources, expose payment instructions, and gather operational intelligence \cite{wood2024ndss,miramirkhani2017ndss,agarwal2025usenix}. Semi-automated platforms have further enabled security teams to scale engagements with scammers~\cite{siadati2020framework,siadati2019prevention}. Prior studies show that scam conversations typically progress through trust-building, urgency creation, and extraction phases \cite{wood2024ndss,edwards2017www,agarwal2025usenix}, with victims complying at non-trivial rates \cite{tu2019users}. While existing work demonstrates the defensive value of sustained engagement \cite{isacenkova2013spw,miramirkhani2017ndss,agarwal2025usenix}, it remains largely unexplored how scammers may adapt their strategies in the presence of scambaiters.

\subsection{Scambaiting Chatbots}
\label{subsec:chatbot_scambaiting}
Automated scambaiting transformed a largely manual defensive practice into a scalable one. Early systems explored conversational agents that could engage scammers over voice or messaging channels for extended periods, demonstrating that automation could sustain attacker attention without continuous human involvement \cite{sahin2017chatbots,siadati2025send,pandit2023combating}. More recent agentic LLM-based systems advance this direction by generating contextually adaptive responses that support longer and less scripted interactions ~\cite{siadati2025send,bajaj2023automatic,Basta2025}. Related work on LLM-based honeypots further suggests that sustaining adversary engagement to extract threat intelligence generalizes beyond scambaiting to broader deception-based defenses~\cite{wang2024honeygpt}. Prior research and industry reports indicate that such systems can simultaneously manage large numbers of fraudulent conversations at scale~\cite{siadati2025send,bajaj2023automatic,Basta2025}.
Existing evaluations primarily measure how effectively these systems act as engaging conversational agents ~\cite{siadati2025send,bajaj2023automatic}. However, the dynamics that emerge when attacker-side responses also become automated, and the resulting need for strategic scambaiting, remain largely unexplored.

\subsection{Game Theory and Strategic Adaptation}
\label{subsec:gt_deception_signals}
Security game literature provides formal foundations for reasoning about adversarial deception and adaptation, including dynamic Bayesian models \cite{huang2019dynamic,huang2021duplicity} and Nash equilibrium analyses of defender investment under heterogeneous attacker populations \cite{grossklags2008secure}. A central insight in cyber deception and security games is that rational decisions depend on belief updates derived from observed behavior and signals \cite{pawlick2019leaky}. These works model strategic interaction under incomplete information, where actions reveal underlying type or intent. However, they differ from the scammer-baiter setting in three important ways. First, existing models such as \cite{huang2019dynamic,huang2021duplicity} treat the hidden type as a single attacker or insider whose identity the defender tries to infer; in scambaiting the uncertainty runs in the opposite direction, with the scammer attempting to infer whether its counterparty is a genuine victim or a defender-controlled agent. Second, signaling game models such as \cite{pawlick2019leaky} treat deception detection as a single-shot or stateless interaction; scammer-baiter engagements progress through a staged kill chain in which operationally sensitive information such as mule account details is only disclosed at late stages, so surviving early screening is a precondition for intelligence capture. Third, none of these models incorporate campaign-level resource attrition, where mule accounts and defender identities are finite pools depleted across repeated interactions and the campaign outcome depends on which side exhausts its resources first. Our work builds on these game-theoretic foundations while addressing these structural differences.

Taken together, prior work establishes three relevant foundations: the defensive value of sustained scammer engagement, the feasibility of automating that engagement at scale, and formal tools for reasoning about adversarial deception under incomplete information. No existing work combines all three in a setting where both sides are automated, scammers screen adaptively on the basis of evolving beliefs, intelligence disclosure is a late-stage outcome that depends on surviving that screening, and mule accounts and defender identities are finite resources depleted across repeated interactions. This paper addresses that gap.

\section{Problem Statement}
\label{sec:problem}
The arms race to adopt agentic technologies on both sides raises the fundamental question: How can agentic scambaiting remain effective against scammers that also employ automated and adaptive systems? In this paper, we study two related problems. First, we study the strategic resource dynamics that emerge when both sides can scale their operations: scammers can reach larger numbers of targets, while agentic scambaiters can engage many scammers simultaneously. In particular, we analyze how this scaling affects the depletion of scammer infrastructure, such as mule accounts and payment channels. Second, we examine whether agentic scambaiters can continue to extract actionable threat intelligence when scammers dynamically adapt their behavior and actively attempt to identify deceptive counterparts.



We articulate five research questions tracing a path from initial deployment through long-term adversarial exposure:

\begin{itemize}
    \item \textbf{RQ1:} How do scammer-side scaling mechanisms, including higher contact throughput, lower operating cost, and wider campaign coverage, change campaign duration, resource depletion, and intelligence yield?

    \item \textbf{RQ2:} How does improved scammer detection of automated defenders affect the tradeoff between aggressive probing and conservative cover-preserving engagement?

    \item \textbf{RQ3:} At what population-level deployment rate do scambaiting agent
    render scamming economically unprofitable, and what constitutes an operating level that balances intelligence yield and attacker deterrence?

    \item \textbf{RQ4:} How do asymmetric improvements by scammers and defenders shape the long-term equilibrium of an emerging arms race, especially with respect to intelligence disclosure rates and scammer utility?

    \item \textbf{RQ5:} How do variations in scammer behavioral strategies, such as persistence versus caution, affect both the quantity of intelligence extracted and the profitability of scamming operations?

\end{itemize}


Analyzing these questions requires moving beyond individual conversations. Scammers lose mule accounts when scambaiters extract payment details, while defender personas lose effectiveness once identified as deceptive. Because both resources are finite, campaign outcomes depend on which side depletes its resources first.

As noted by Herley, scammers act as rational economic agents who avoid unprofitable targets~\cite{herley2012weis}. The key question is therefore not whether a scambaiter can sustain a conversation, but whether it can succeed against an adversary that continuously updates its beliefs from conversational signals. Existing scambaiting frameworks do not fully capture these adaptive dynamics. To address this gap, we develop a model that enables systematic analysis of belief evolution, strategic adaptation, and utility trade-offs, with the scammer's belief about whether the counterparty is a victim or a scambaiter serving as the central state variable.

\section{Foundation for Strategic Scambaiting}
\label{sec:foundation}

Online fraud is a competition over finite operational resources. Scammers depend on mule accounts, cryptocurrency wallets, and communication channels to monetize campaigns. Once these assets are identified and reported, they must be replaced at cost, making disclosure a direct operational loss. Defenders similarly rely on limited resources, including identities and communication accounts used by scambaiting agents. When a persona is recognized as deceptive, it may lose effectiveness and require replacement. Because both sides incur costs to replenish resources, campaign outcomes depend on long-term resource attrition rather than individual conversations. This work focuses on the engagement phase after initial contact, excluding upstream filtering and platform-level interventions.

These constraints make scambaiting a strategic contest in which both parties adapt under uncertainty. To analyze this interaction, we develop a framework that models conversational signals, belief updates, and capability differences between scammers and defenders. The framework further captures resource attrition, incentives, and security-relevant outcomes, enabling systematic analysis of campaign-level dynamics.




\subsection{Scam Engagement Structure}
\label{subsec:engagement_structure}

Scambaiting represents a unique interaction between defenders and scammers that requires modeling both parties' behavior and decision-making. This subsection describes the scammer behavior during the scamming phase, the staged progression through which scam engagements evolve, and the information available to each side. Throughout the paper, \emph{baiter}, \emph{scambaiter}, \emph{baiting agent}, \emph{agent}, and \emph{defender} all refer interchangeably to the defender-controlled persona, while \emph{scammer} and \emph{attacker} both refer to the scammer.

\paragraph{Target Screening}
A scammer begins a conversation without knowing whether the 
counterparty is a genuine victim or a defender-controlled baiting 
agent. In the framework, the scammer faces a single opponent whose 
type is hidden (Appendix~\ref{app:hidden_type}). With probability 
$\rho$ the counterparty is a strategically operated baiting agent 
working to extract operational intelligence. With probability 
$1{-}\rho$ the counterparty is a genuine victim following 
behavioral patterns shaped by fear, confusion, or compliance 
rather than strategic reasoning. The baiting agent actively 
chooses responses to maximize the chance of reaching late-stage 
disclosure. The genuine victim has no awareness that the 
interaction is a scam and no objective to optimize. Because the 
scammer cannot observe which type it faces, every decision in the 
conversation must account for both possibilities simultaneously.

\begin{figure}[t]
    \centering
    \includegraphics[width=0.85\linewidth]{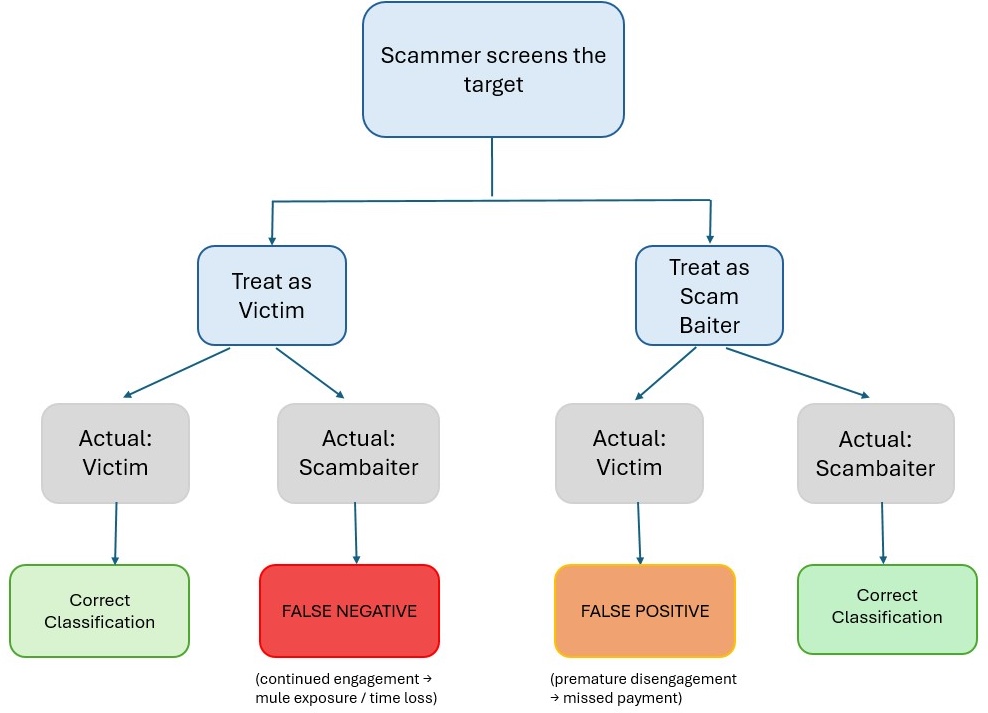}
    \caption{Scammer Target Screening Errors}
    \label{fig:asymmetric-errors}
\end{figure}

This uncertainty is operationally consequential because the two 
types of error the scammer can make carry asymmetric costs. 
As shown in Figure~\ref{fig:asymmetric-errors}, a false positive 
occurs when the scammer treats a real victim as a baiting agent and 
disengages early. The cost is a missed payment opportunity, but 
no infrastructure is exposed. A false negative occurs when the 
scammer treats a baiting agent as a genuine victim and escalates 
far enough to reveal payment endpoints such as mule accounts. The 
cost is not only wasted engagement time but potential exposure of 
fragile infrastructure that defenders can force into retirement. 
This asymmetry is what makes the screening problem non-trivial. 
Moving quickly toward a payment request is profitable when the 
counterparty is a genuine victim. The same behavior is costly when 
the counterparty is a baiting agent because it commits operational 
resources to an engagement that cannot pay out and may be designed 
to extract infrastructure details before the scammer disengages. 
Slowing down to screen more carefully reduces the risk of a false 
negative but delays progress with real victims and can reduce 
payment success. Because the counterparty type is never directly 
observable, the scammer must resolve this dilemma on the basis of 
accumulated conversational evidence alone.

\begin{figure}[t]
\centering
\includegraphics[width=\linewidth]{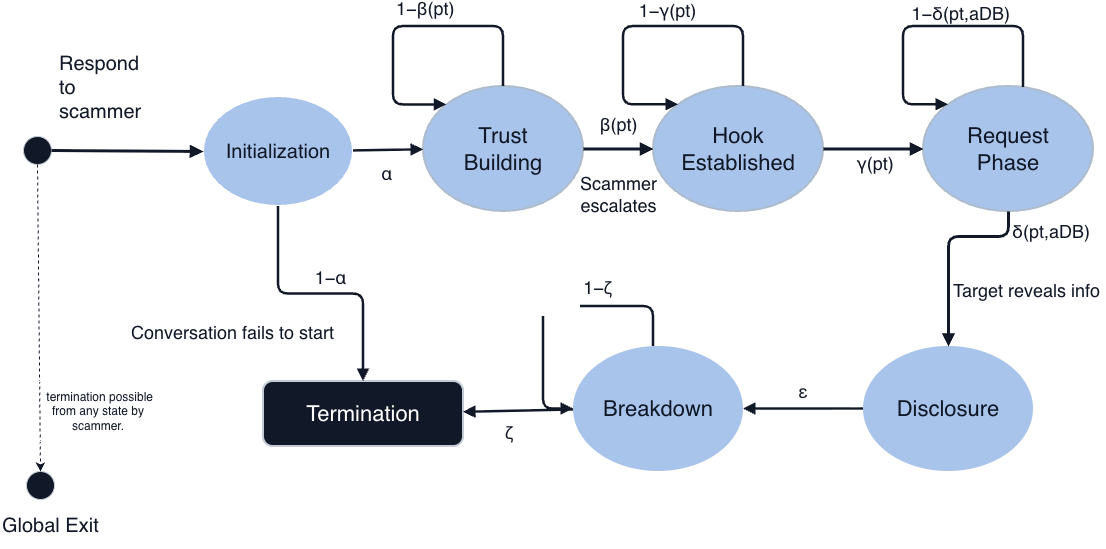}
\caption{Scam engagement kill chain. Each stage represents a 
distinct operational phase with different objectives and 
available actions for both sides. Stage advancement is 
belief-driven.}
\label{fig:kill_chain}
\end{figure}

\paragraph{Scam Engagement Kill Chain}
Scam conversations typically progress through recognizable phases rather than random exchanges. The defensive value of scambaiting depends on advancing far enough into the interaction to elicit operational details such as mule accounts and payment instructions. Accordingly, we model engagement as a kill chain in which progression is driven by the scammer's confidence that the target is worth further investment (Figure~\ref{fig:kill_chain}).

\emph{Trust Building} establishes credibility, \emph{Hook Established} introduces the scam narrative, and \emph{Request Phase} seeks payment or credentials. \emph{Disclosure} is the primary objective of the baiting agent, as it may reveal payment channels and other intelligence artifacts. \emph{Breakdown} captures interactions in which suspicion emerges but communication continues, while \emph{Termination} ends the episode. Progression is neither scripted nor time-based. Scammers advance, probe, stall, or abandon the interaction according to their evolving beliefs about the target. The formal transition model is presented in Appendix~\ref{app:transition_model}, and the corresponding action spaces are given in Table~\ref{tab:action_spaces}.

\paragraph{Information Constraints}
The two sides operate under different informational constraints. The scammer's strategy conditions on the current stage and its running belief about counterparty type (Appendix~\ref{app:strategies}). The baiting agent's strategy conditions on the stage and the scammer's most recent action, but not on the scammer's belief, so it must steer toward Disclosure without observing whether its responses are shifting the scammer's assessment. The genuine victim is non-strategic and responds based on fear, compliance, or confusion.

\subsection{Belief, Signals, and Adaptation}
\label{subsec:signals_adaptation}

The screening problem established in \S\ref{subsec:engagement_structure} 
depends on a mechanism by which the scammer learns about the counterparty 
type as the conversation unfolds. This subsection specifies that mechanism 
and then introduces the two parameters that govern how much each side can 
shift it in their favor.

\paragraph{Conversational signals and belief updating}
At each turn the interaction produces a coarse conversational signal 
$\sigma_t \in \{\text{Green}, \text{Yellow}, \text{Red}\}$, where 
Green denotes behavior consistent with a genuine victim, Yellow is 
ambiguous, and Red raises suspicion of a baiting agent. The signal is 
generated by the defender's action and actual type, not directly observed 
by the scammer as a labeled category but inferred from the conversational cues produced during the exchange. The scammer uses each observed signal to update its 
confidence that the counterparty is a genuine victim via Bayes' rule:
\begin{equation}
p_{t+1}
=
\frac{P(\sigma_t \mid \text{Victim})\,p_t}
     {P(\sigma_t \mid \text{Victim})\,p_t
      + P(\sigma_t \mid \text{Baiter})\,(1-p_t)},
\label{eq:bayes_update_main}
\end{equation}
where $p_t$ is the scammer's current belief that the counterparty 
is a genuine victim and $P(\sigma_t \mid \cdot)$ are the type-conditional 
signal likelihoods. The baseline values for these likelihoods are given in 
Appendix~\ref{app:signal_model}.


The two types produce overlapping signal distributions, with victims more likely to produce Green signals and baiters more likely to produce Yellow or Red signals, so no single observation resolves the scammer's uncertainty. The overlap is intentionally kept large enough that signals are only weakly informative per turn, placing the model in a regime where sustained engagement is both necessary and uncertain. This produces gradual belief movement over multiple turns, consistent with the empirical observation that scammers sustain engagements for many exchanges before abandoning. The scammer's prior at the start of each engagement is $p_0 = 1 - \rho$, where $\rho$ is the population-level prevalence of baiting agents it expects to encounter. Action-category modifiers that shift these base likelihoods depending on what the defender actually does are detailed in Appendix~\ref{app:signal_model}, along with the full belief update specification.

\paragraph{Scammer detection capability ($\gamma$)}
As AI-driven scam operations mature, scammer tooling become 
better at extracting distinguishing behavioral cues from 
multi-turn exchanges regardless of what either side chooses to 
do. A sophisticated scammer platform might analyze 
response timing, linguistic patterns, or consistency across 
turns in ways that pull the signal distributions for genuine 
victims and baiting agents further apart. The parameter 
$\gamma \in [0, 1]$ captures this possibility by controlling 
how well the scammer can distinguish baiters from genuine victims. At $\gamma = 
0$ the scammer cannot distinguish a baiter from a genuine victim.
As $\gamma$ increases, genuine victims move toward 
producing only Green signals and baiting agents move toward 
producing only Red signals, until at $\gamma = 1$ each type 
emits a perfectly diagnostic signal and the scammer can 
identify the counterparty type with certainty after a single 
turn. Critically, this is a change in what both types actually 
produce, not in how the scammer processes the evidence. The 
scammer benefits automatically because the signals it observes 
carry more information about type without the scammer having 
to do anything differently. 

\paragraph{Baiter mimicry quality ($\beta$)}
A defender deploying a scambaiting system can invest in making 
the baiting agent's conversational behavior look more like 
that of a genuine victim, for example by training on real 
victim transcripts, tuning response patterns, or reducing 
the probing behaviors that generate Red signals. The parameter 
$\beta \in [0, 1]$ captures the effect of that investment. At 
$\beta = 0$ the baiting agent produces the baseline signal 
distribution, which is noticeably different from a genuine 
victim's across all three categories. As $\beta$ increases, the baiting agent's signal distribution shifts toward the victim distribution, making 
the two types progressively harder to distinguish from the 
scammer's perspective. At $\beta = 1$ the baiting agent's 
signals are statistically identical to those of a genuine 
victim, eliminating all signal-level evidence the scammer 
could use to identify it. Genuine victim behavior is 
unaffected by $\beta$ throughout, so this parameter isolates 
the effect of defender investment in persona quality without 
changing anything about how the scammer reasons or what 
genuine victims produce. A scammer's detection capability is therefore bounded by the defender's mimicry quality: higher $\gamma$ provides 
diminishing returns as $\beta$ increases, since detection 
operates on signals that have already been made to resemble 
a genuine victim's.

\subsection{Incentive Structure and Engagement Outcomes}
\label{subsec:incentives}

Our kill chain and signal model define the structure and information flow of the interactions. What drives the decisions within that structure 
is the incentive each side faces at every turn. This 
subsection specifies those incentives, the terminal outcomes 
they produce, and how outcomes accumulate into campaign-level 
resource attrition across repeated interactions.

\paragraph{Per-turn and terminal payoffs}
The scammer pays a per-turn engagement cost $c_b$ when interacting with a baiting agent and $c_v = 0$ against a genuine victim. Its expected per-turn cost at belief $p_t$ is $c_b(1-p_t) + c_v \cdot p_t$, which simplifies to $c_b(1-p_t)$ since $c_v = 0$. The baiting agent earns a per-turn occupancy reward $r_{\text{occ}}$ for keeping the scammer engaged. If the scammer abandons, it incurs a penalty $p_{\text{ab}}$ with no terminal payoff to the defender. If the engagement reaches Disclosure or Breakdown and the scammer takes a commit action such as sharing mule account numbers or cryptocurrency wallet addresses to proceed toward payment, the defender may capture that infrastructure intelligence yielding $r_{\text{intel}}$, with capture probability depending on the stage and whether the baiter actively probes. Against a genuine victim, the same commit action allows the scammer to extract payment yielding $r_{\text{pay}}$. Table~\ref{tab:payoff_main} maps each event to its parameter 
name and instantiated value. The full episode-level utility  specification is in Appendix~\ref{app:utilities}.

\begin{table}[t]
\centering
\caption{Payoff events, symbols, and values}
\label{tab:payoff_main}
\scriptsize
\begin{tabular}{l|c|r}
\hline
\textbf{Payoff event} & \textbf{Symbol} & \textbf{Value} \\
\hline
Defender per turn (engagement) & $r_{\text{occ}}$   & $+3$   \\
Scammer per turn vs.\ agent    & $c_b$              & $-5$   \\
Scammer per turn vs.\ victim   & $c_v$  & $0$    \\
Defender intel capture         & $r_{\text{intel}}$ & $+150$ \\
Scammer payment extraction     & $r_{\text{pay}}$   & $+100$ \\
Scammer abandonment penalty    & $p_{\text{ab}}$    & $-20$  \\
\hline
\end{tabular}
\end{table}

\paragraph{Generalizability of the payoff structure}
The model's qualitative findings do not depend on these  specific values but on four conditions the instantiation satisfies. The first is $r_{\text{intel}} > r_{\text{pay}} > 0$, meaning intel capture is worth more than payment extraction, which makes Disclosure the primary defensive objective. The second is $c_b < 0$, meaning each turn spent with a baiting agent is a net loss for the scammer, which is the asymmetry that makes screening rational. The third is $r_{\text{intel}} > |p_{\text{ab}}|$, meaning capturing intel is worth more than the abandonment penalty, so early exit is not the scammer's dominant strategy. The fourth is 
$r_{\text{intel}} > |c_b| \cdot \bar{T}$ where $\bar{T}$ is the mean turns to disclosure, meaning pursuing engagement is economically rational even after accounting for per-turn costs. Any parameterization satisfying these four conditions produces the same qualitative equilibrium structure.

\paragraph{Campaign-level resource depletion}
A single episode captures only part of the operational 
picture. Scammers maintain a finite pool of mule accounts, 
and defenders maintain a finite pool of email identities 
through which baiting agents engage. When a baiting agent 
successfully captures intel in an episode, the exposed mule 
account must be retired from the scammer's pool. When an 
episode terminates from Breakdown with a baiting agent as 
the counterparty, the scammer has identified the engagement 
as a defensive operation and the email identity is burned 
from the defender's pool. Neither pool replenishes 
automatically. The campaign begins with 1,000 defender email 
identities and 50 scammer mule accounts, and terminates 
when either pool is exhausted. This attrition structure 
connects directly to the resource competition introduced in the beginning of 
\S\ref{sec:foundation}: the payoffs within a single 
episode determine how quickly each side depletes the other's 
operational capacity across a campaign.

\paragraph{Security-relevant outcome metrics}
The framework produces four metrics that connect episode 
outcomes to the research questions in Section~3. The 
\emph{intelligence disclosure rate} (IDR) is the fraction 
of baiter episodes in which intel capture succeeds. It 
measures whether the defense is generating actionable 
operational intelligence or merely occupying scammer time. 
The \emph{payment rate} is the fraction of victim episodes 
in which the scammer extracts payment. It measures the 
harm reaching genuine victims under the deployed defense. 
The \emph{abandonment rate} is the fraction of all episodes 
that end in the scammer choosing to disengage. It reflects 
how aggressively the scammer is screening and how much 
deterrence the defense imposes. \emph{Mean turns per 
episode} measures how long engagements last on average, 
capturing the time cost the defense imposes on scammer 
operations regardless of whether disclosure is reached. 
These four metrics are reported for every experimental 
condition in Section~7.

\section{Computational Approach for Strategic Engagement}
\label{sec:formal_model}

\subsection{Computing Optimal Engagement Strategies}
\label{subsec:computing_strategies}

Finding how each side should act in this interaction is not tractable
analytically. Three properties of the problem together make closed-form
solution unavailable. The scammer's belief about defender type is a
continuous value that evolves differently for every possible sequence
of signals, so evaluating any strategy's expected value requires
accounting for all possible belief trajectories over the full 25-turn
horizon. The scammer also optimizes against a mixture of two distinct
transition dynamics, one applying when the counterpart is an agent and one
when a genuine victim, weighted by the current belief, and which dynamic
applies is unknown at decision time. The baiter cannot observe the scammer's current belief and must choose actions that perform
well across the full range of beliefs the scammer might hold, weighting
each belief by the probability the scammer actually holds it at that
point in the interaction.

We handle the continuous belief problem by discretizing the belief space
into ten bins at $\mathcal{B} = \{0.05,\, 0.15,\, \ldots,\, 0.95\}$,
reducing the infinite-dimensional space to a finite grid without
materially altering the model's qualitative behavior
(Appendix~\ref{app:policy_computation}). For each player we then compute
a \emph{value function} over the product space of kill-chain stages,
discretized beliefs, and turns remaining, working backwards from $t = 0$
to $t = 25$ via \emph{backward induction}.

For the scammer, the value $V_{\mathrm{SB}}(s,\, b,\, t)$ is the
expected payoff from state $s$ at belief $b$ with $t$ turns remaining
given the current baiter strategy. At each step the scammer selects the
action that maximizes expected payoff, where the continuation value at
each bin is the signal-weighted sum of future values after applying the
Bayesian update across the mixture of baiter and victim dynamics at the
current belief. The scammer's optimization is over stage-level actions:
whether to escalate the scam narrative, probe for inconsistencies, or
abandon the target.

For the baiter, the value function runs over the product space of stages
and observed scammer actions rather than stages and beliefs, because the
baiter conditions on what it can observe. The baiter's value at each
\textup{(stage, scammer action)} pair is averaged over all belief bins
the scammer might hold at that point, weighted by how probable the
observed scammer action is at each belief level under the scammer's
current strategy. This weighting is what makes the baiter's computation
depend on the scammer strategy: a baiter response is only meaningful
when evaluated against the distribution of belief states in which the
scammer would actually take that action
(Appendix~\ref{app:policy_computation}).

\paragraph{Inference sharpness ($\alpha$)}
A third parameter, $\alpha$, controls how strongly the scammer's belief
update responds to each observed signal during the solve. At $\alpha = 1$
the scammer applies standard Bayes' rule. At $\alpha > 1$ signals are
treated as more diagnostic than their empirical frequencies imply,
accelerating belief movement toward one type or the other, while at
$\alpha < 1$ updates are more sluggish. Unlike $\beta$ and $\gamma$,
which change what signals the environment generates, $\alpha$ changes
only how the scammer processes those signals. It therefore shifts
equilibrium behavior without altering the information structure itself.
The update rule incorporating $\alpha$ is given in
Appendix~\ref{app:signal_model}.

\textsc{ComputeEquilibrium} (Algorithm~\ref{alg:equilibrium})
orchestrates this computation by alternating the two backward-induction
solves until neither player can improve against the other's current
strategy. Each iteration applies a damped update with factor
$\eta = 0.2$, mixing the new best response with the current strategy
rather than replacing it outright, which prevents the oscillation that
arises when one player's improvement destabilizes the other's solution. Convergence is
declared when the total $\ell_1$ distance between successive strategies
falls below $\varepsilon = 10^{-4}$. A 2-cycle detector resolves any
remaining oscillation by averaging the two most recent strategy profiles
and halting. Across all parameter configurations reported in this paper
the procedure converges in under 50 iterations.

\begin{algorithm}[t]
\caption{\textsc{ComputeEquilibrium}: Damped Iterative Best Response}
\label{alg:equilibrium}
\textbf{Require:} Damping $\eta = 0.2$;
convergence threshold $\varepsilon = 10^{-4}$;
cycle tolerance $\varepsilon_{\mathrm{cyc}} = 10^{-6}$;
max iterations $K = 200$
\begin{algorithmic}[1]
\State Initialize $\pi_{\mathrm{SB}}$ and $\pi_{\mathrm{B}}$ as uniform strategies
       over all state--belief pairs
\For{iteration $k = 1, 2, \ldots, K$}
    \State Compute scammer best response $\hat{\pi}_{\mathrm{SB}}$ via
           backward induction against current $\pi_{\mathrm{B}}$
    \State Compute baiter best response $\hat{\pi}_{\mathrm{B}}$ via
           backward induction against current $\pi_{\mathrm{SB}}$
    \State $\pi_{\mathrm{SB}} \leftarrow (1{-}\eta)\,\pi_{\mathrm{SB}} + \eta\,\hat{\pi}_{\mathrm{SB}}$
    \State $\pi_{\mathrm{B}}   \leftarrow (1{-}\eta)\,\pi_{\mathrm{B}}   + \eta\,\hat{\pi}_{\mathrm{B}}$
    \If{total $\ell_1$ movement since last step $< \varepsilon$}
        \State \Return $(\pi_{\mathrm{SB}},\, \pi_{\mathrm{B}})$
    \EndIf
    \If{$k \geq 3$ \textbf{and} strategies match those from two steps ago within $\varepsilon_{\mathrm{cyc}}$}
        \State Resolve 2-cycle: \Return average of the two most recent
               strategy profiles
    \EndIf
\EndFor
\State \Return $(\pi_{\mathrm{SB}},\, \pi_{\mathrm{B}})$
\end{algorithmic}
\end{algorithm}



\subsection{From Strategies to Observed Outcomes}
\label{subsec:simulation_pipeline}

Solving for stable strategies answers what each side does at every
state and belief. It does not answer how often those strategies produce
a given outcome. These quantities cannot be read off the strategy
tables because the interactions that produce them are stochastic.
Signals are random draws from a type-dependent distribution, state
transitions are sampled from distributions that depend on the current
state and actions taken, and terminal outcomes are probabilistic
conditional on reaching the relevant stage. Averaging over these
random draws requires a \emph{solve-then-simulate} pipeline, in which
stable strategies are first computed and then used to drive many
independent interactions whose outcomes are aggregated.

\paragraph{The belief update asymmetry}
One structural property of the simulator shapes the belief update at
every turn. The signal the scammer observes is generated from the
defender's \emph{realized action}, but the belief update uses the
\emph{policy-implied action} rather than the realized one, which is the correct treatment because the scammer observes only the conversational signal, not what the defender actually did to produce it.

For \emph{campaign}-level results we run the same episode simulator
repeatedly without resetting resource pools between episodes. After
each episode the pools are updated following the depletion rules from
Section~\ref{sec:foundation}. An agent episode that terminates from
Breakdown burns one email identity, and any episode in which intel is
captured burns one mule account. The campaign ends when either pool
reaches zero.


\section{Experiment}
\label{sec:simulation}

\subsection{Empirical Basis}
\label{subsec:empirical-basis}
The model structure is informed by empirical observations from prior work evaluating an LLM-based scambaiting system in real-world scam engagements~\cite{siadati2025send}.The dataset contains over 2,600 engagements and 18,700 messages, offering insight into engagement dynamics, disclosure timing, and abandonment behavior. These observations motivate the model structure and inform the choice of a plausible baseline regime, but are not used to statistically estimate the full parameter set.




\subsection{Simulation Setup}
\label{subsec:sim_setup}
We implemented the model in Python, with each model variable represented as a configurable parameter. For a given parameter setting, equilibrium strategies are first computed using the procedure described in Section~\ref{subsec:computing_strategies}. These strategies are then held fixed while executing (N=2000) independent simulation episodes. Performance metrics are estimated as sample averages across all episodes. The simulation parameters and evaluation metrics, including intelligence disclosure rate (IDR), payment rate, abandonment rate, and mean conversation length, are summarized in Table~\ref{tab:sim_params} in the Appendix.

\subsection{Baseline Operating Point}
\label{subsec:baseline}

We calibrated two cost parameters, the scammer's per-turn engagement cost $c_b$ and the abandonment penalty $p_{\text{ab}}$, against three empirical quantities reported in deployment data, an IDR of approximately 31.7\% among matured engagements, a mean of 10.3 turns to disclosure, and a non-negligible abandonment rate. The best-fitting configuration sets $c_b = -5$ and $p_{\text{ab}} = -20$, yielding a baseline IDR of 22.9\% and a mean of 9.89 turns to disclosure. This gap is intentional. Configurations that closely match the 31.7\% figure produce near-zero abandonment, which is inconsistent with observed scammer behavior, and the two figures are not directly comparable since the empirical value is conditional on matured engagements while the simulation IDR is unconditional over all agent episodes. The initial engagement probability was set to match the observed contact acceptance rate of 48.7\%, and all remaining transition parameters are modeling assumptions rather than fitted quantities, so abandonment and payment-extraction rates are emergent outputs of the calibrated cost structure. Sensitivity analysis confirms the qualitative findings are robust. Across all signal configurations, IDR decreases as attacker screening capability ($\gamma_{\text{detection}}$) increases, while improvements in agent mimicry ($\beta_{\text{mimicry}} > 0$) partially mitigate this effect.

The two terminal reward values $r_{\text{intel}} = 150$ and $r_{\text{pay}} = 100$ satisfy the four ordinal conditions in Section~\ref{subsec:incentives}. At $c_b = -5$ and the empirically observed $\bar{T} = 10.3$ turns to disclosure~\cite{siadati2025send}, the total engagement cost of reaching disclosure is $|c_b| \cdot \bar{T} = 51.5$, so $r_{\text{intel}} = 150$ yields a net return of $+98.5$ per successful capture, making pursuit of disclosure economically rational under the calibrated cost structure. The lower value of $r_{\text{pay}}$ reflects that a single victim payment is a one-transaction gain, whereas mule account intelligence carries operational value beyond the episode in which it is captured, since it can be shared with financial institutions to disable the exposed infrastructure~\cite{siadati2025send}.

\begin{table}[t]
\centering
\caption{Baseline operating point summary.}
\label{tab:baseline_summary}
\scriptsize
\begin{tabular}{l r}
\hline
Metric (baseline, $\rho = 0.30$) & Value \\
\hline
Agent fraction ($\hat{\rho}$) & 29.2\% \\
IDR (intel rate $\,|\,\text{Bot}$) & 22.9\% \\
Payment rate ($\,|\,\text{Victim}$) & 53.4\% \\
Abandon rate & 14.8\% \\
Mean turns & 9.89 \\
Mean scammer payoff (overall) & 28.03 \\
Mean scammer payoff ($\,|\,\text{Bot}$) & $-32.33$ \\
Mean scammer payoff ($\,|\,\text{Victim}$) & 52.92 \\
\hline
\end{tabular}
\end{table}

Campaign runs additionally track resource consumption on both sides following the depletion rules in Section~\ref{subsec:incentives}, complementing the per-episode payoffs reported in Table~\ref{tab:baseline_summary}.


\section{Model Evaluation and Analysis}
\label{sec:results}

Using the defined framework, we analyze how changes in key parameters affect scambaiting results including intelligence collection and campaign-level outcomes. The objective is not to predict real-world deployment performance, but rather to understand the strategic dynamics implied by the model and identify the conditions under which qualitative behavioral changes occur.
Accordingly, quantities such as break-even points, tipping points, and operational thresholds should be interpreted as properties of the baseline model specification. They indicate parameter regimes where equilibrium behavior changes direction (e.g., from engagement to abandonment or from profitable to unprofitable operation), rather than precise forecasts of real-world outcomes. The value of these results lies in revealing the relative effects of different capabilities and incentives, and in identifying the mechanisms that drive strategic adaptation on both sides. The results are organized into five subsections, each addressing a distinct research question introduced in Section~\ref{sec:problem}.

\subsection{Scammer Earnings Turn Negative Above a Scambaiter Density Threshold}
\label{sec:results_threshold}

This subsection examines how the proportion of scambaiting agents in the target population affects scammer behavior, profitability, and intelligence collection. While it may seem that increasing the fraction of scambaiting agents would reduce scammer profitability, the relationship is not straightforward: scambaiting campaigns themselves consume resources, and scammers are not passive adversaries. As scambaiter prevalence increases, rational scammers update their beliefs about whether a target is genuine and adjust their engagement strategy accordingly, which may reduce scammer revenue, increase abandonment, or discourage engagement altogether. Although reduced scam activity is desirable from a defensive perspective, it may also decrease opportunities for intelligence collection.

Each conversation a scammer starts carries an operating cost paid turn by turn, whether the counterpart is a genuine victim who may eventually send money or a scambaiter that never will. The scambaiter prevalence $\rho$ therefore sets the fraction of the scammer's effort that cannot produce a payout, and also sets the scammer's prior belief entering each new conversation: assuming the scammer knows the true prevalence, the prior probability of facing a genuine victim is $1 - \rho$, so a scammer facing high $\rho$ begins every contact already suspecting a scambaiter and abandons quickly, while one facing low $\rho$ stays engaged. Because belief matches reality in this setup, the behavioral changes across the sweep are sharp. We sweep $\rho$ from $0.05$ to $0.80$, recomputing the scammer's best response at each value.

\begin{figure}[t]
  \centering
  \includegraphics[width=\linewidth]{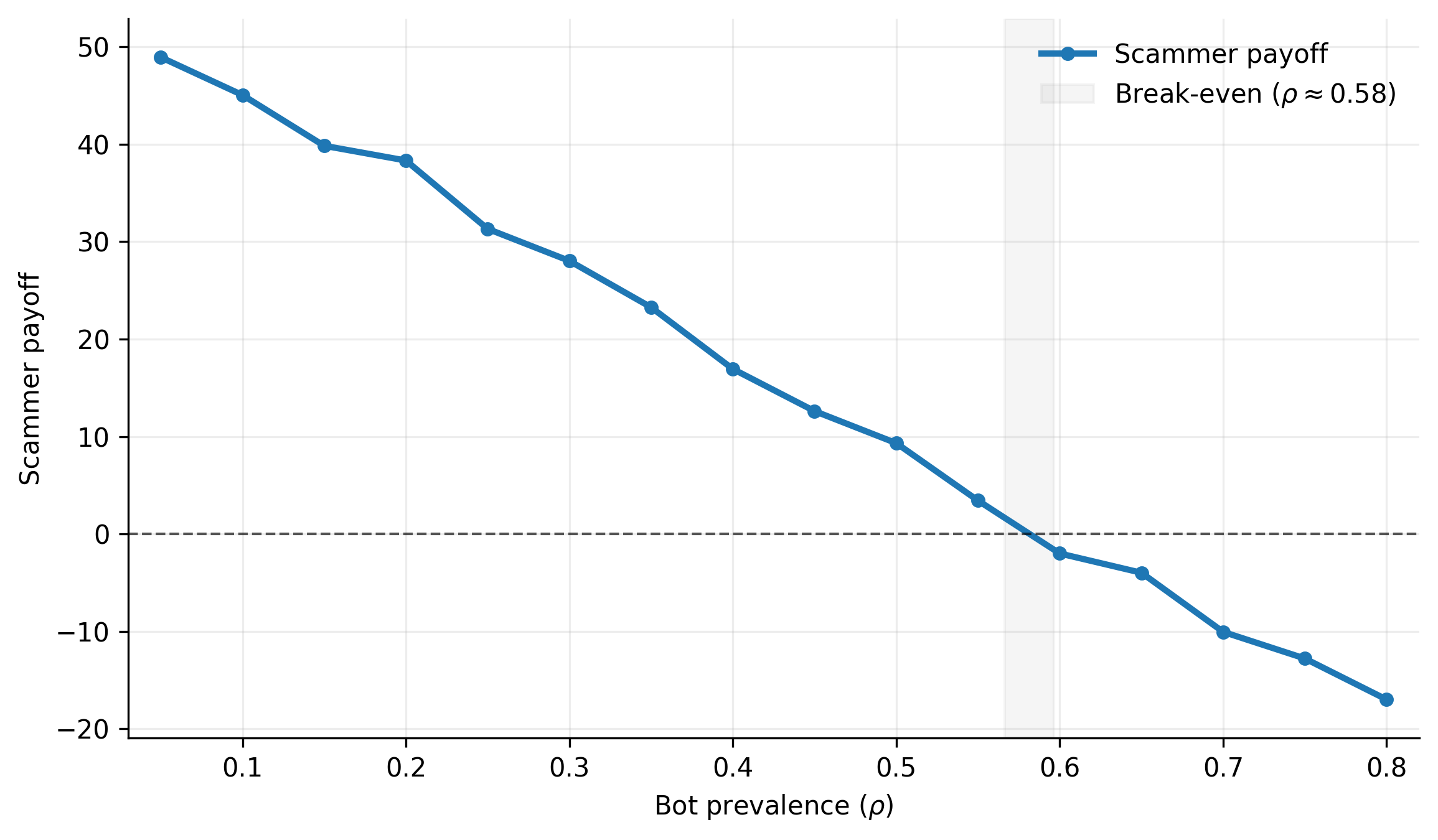}
  \caption{Mean scammer payoff per episode as scambaiter prevalence $\rho$
  increases. The shaded band marks the break-even region where payoff
  changes sign under our baseline parameters.}
  \label{fig:profitability_vs_rho}
\end{figure}

Figure~\ref{fig:profitability_vs_rho} shows scammer earnings per episode
falling steadily with $\rho$ and changing sign between $\rho=0.55$ and
$\rho=0.60$, the point above which earnings from victim conversations
no longer cover the losses absorbed on scambaiter conversations and the
average scamming operation runs at a loss.

\begin{figure}[t]
  \centering
  \includegraphics[width=\linewidth]{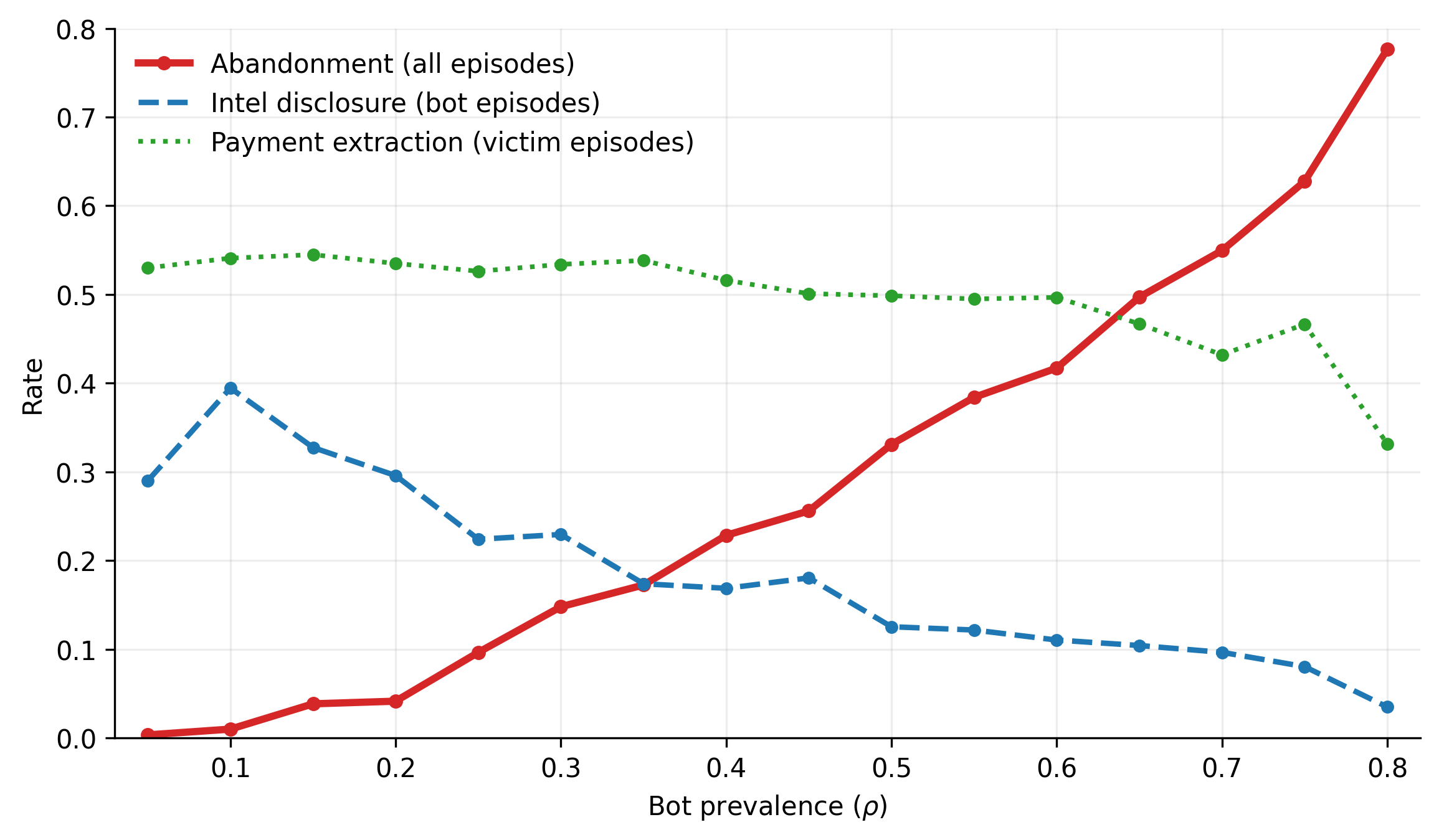}
  \caption{Three engagement outcomes across the scambaiter -prevalence sweep.
  Abandonment and intelligence disclosure are measured on all episodes
  and scambaiter episodes respectively. Payment extraction is conditional on
  victim episodes.}
  \label{fig:outcomes_vs_rho}
\end{figure}

Figure~\ref{fig:outcomes_vs_rho} shows where the lost earnings come from
by tracking three rates across the same sweep. As scambaiter prevalence rises,
abandonment climbs steeply (from under $1\%$ at $\rho=0.05$ to $77.6\%$
at $\rho=0.80$) while intelligence disclosure falls ($39.4\%$ to $3.5\%$),
as the scammer cuts agent-looking conversations short to limit losses,
which also shortens the window in which an agent can extract mule account
details. Payment extraction, the green line, tells a different story:
conditional on reaching a real victim, the scammer's success rate holds
near $50\%$ across most of the sweep and only drops at very high $\rho$,
so the scammer is not getting worse at monetizing victims, simply
encountering fewer of them.

\begin{figure}[t]
  \centering
  \includegraphics[width=\linewidth]{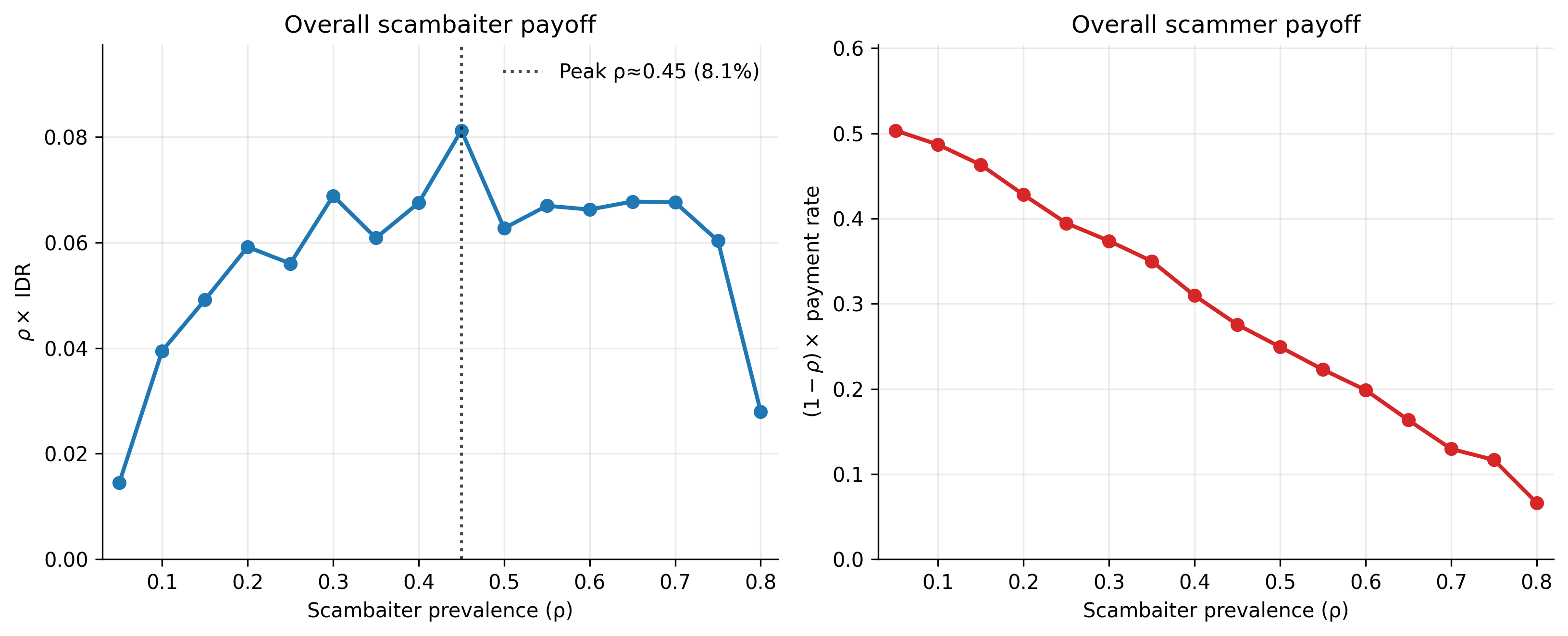}
  \caption{Aggregate payoffs across the scambaiter-prevalence sweep.
  The left panel shows overall scambaiter payoff ($\rho \times \text{IDR}$),
  peaking near $\rho = 0.45$. The right panel shows overall scammer payoff
  ($(1-\rho) \times \text{payment rate}$), declining steadily as fewer
  victim contacts remain.}
  \label{fig:aggregate_payoffs}
\end{figure}

Per-scambaiter disclosure rate declines monotonically with $\rho$, but total intelligence collected per 1000 contacts peaks near $\rho \approx 0.45$ at roughly 81 disclosures before falling: at $\rho = 0.10$ only 39 disclosures occur despite each scambaiter being individually productive, simply because there are too few of them, while at $\rho = 0.80$ scambaiters are numerous but the scammer abandons so aggressively that almost none reach disclosure. Economic disruption and intelligence collection therefore optimize at different deployment scales, with the scammer's operation turning unprofitable near $\rho \approx 0.58$ but peak intelligence yield achieved near $\rho \approx 0.45$, so a deployment targeting maximum disruption requires a higher scambaiter share than one targeting maximum intelligence collection. Figure~\ref{fig:aggregate_payoffs} shows both curves directly.

The exact threshold location depends on the model's payoff and signal parameters, but the sweep shows that such a threshold exists and that the gap between operating below it and above it is large.
\subsection{Probing Extracts More Intel but Accelerates Detection}
\label{sec:results_detection}

The second pattern examines what happens to scambaiting outcomes as scammers get better at reading the conversation. Every scam conversation passes through the kill chain phases established in Section~\ref{subsec:engagement_structure}. By the time it reaches the Request phase, the scambaiter faces a decision. It can push early for mule account details, accelerating toward Disclosure before the scammer grows suspicious. Or it can hold cover, behave like a genuine victim, and let the conversation advance at the scammer's pace. The first approach extracts intelligence faster but emits stronger signals that something is wrong. The second approach preserves cover but surrenders the initiative. We model it as two postures and ask which serves defenders better as scammers improve their ability to read conversational signals. That ability is captured by $\gamma$, a parameter controlling how clearly scammer and victim messages differ, ranging from $\gamma=0$, where a scambaiter's messages are indistinguishable from a real victim's, to $\gamma=1$, where the scammer can identify the difference from cues alone. We sweep $\gamma$ from $0$ to $1$, fixing each posture's behavior and letting the scammer adapt at each value.

\begin{figure*}[t]
  \centering
  \includegraphics[width=\textwidth]{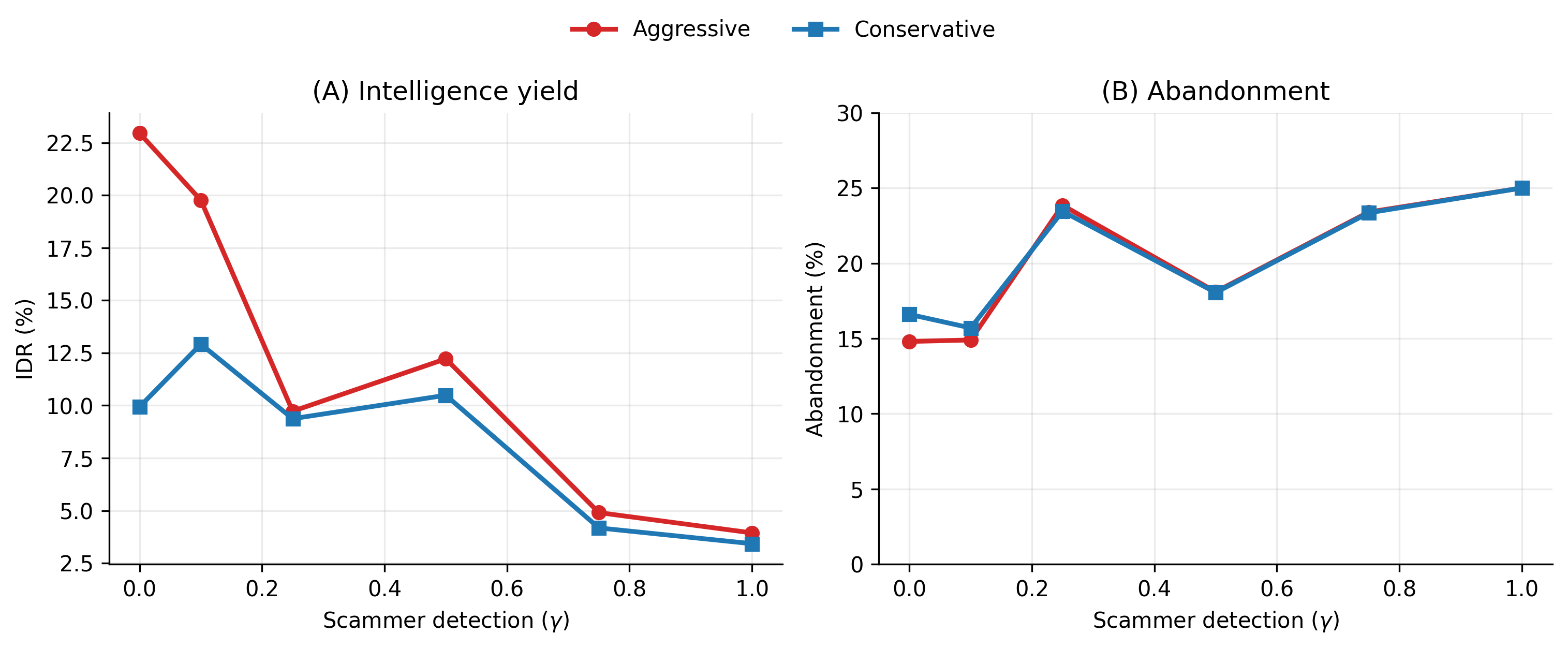}
  \caption{Intelligence disclosure and abandonment rates as scammer detection capability $\gamma$ rises. The two postures converge as $\gamma \to 1$.}
  \label{fig:rq2_detection}
\end{figure*}

Figure~\ref{fig:rq2_detection} shows both postures' intelligence yield and abandonment across the sweep. At $\gamma=0$ the aggressive scambaiter extracts intelligence from $22.9\%$ of scambaiter conversations against the conservative scambaiter's $9.9\%$. The aggressive posture starts more than twice as productive. By $\gamma=0.25$ that gap has closed to $9.7\%$ versus $9.4\%$. By $\gamma=1.0$ both IDR and abandonment converge to nearly identical values across both postures.

The relationship between $\gamma$ and IDR is not linear. For the aggressive posture, IDR falls sharply in the first quarter of the range, dropping from $22.9\%$ at $\gamma=0$ to $9.7\%$ at $\gamma=0.25$, a reduction of more than half. The decline then slows, reaching $3.9\%$ at $\gamma=1.0$. The conservative posture follows a different pattern. Starting at $9.9\%$, it declines more gradually across the full range, ending at $3.4\%$. The steepness of the aggressive curve at low $\gamma$ reflects how much of its advantage depends on the scammer being unable to read probing signals. Once that ability crosses even a modest threshold, the aggressive edge collapses quickly.

The aggressive scambaiter's early advantage comes from probing. At $\gamma=0$ the aggressive scambaiter spends $12.8\%$ of agent-conversation turns on probe actions that push for account details, while the conservative agent spends none. Probing extracts intelligence faster when the scammer cannot tell it apart from a real victim's questions. As $\gamma$ rises, the same probing behavior becomes the clearest cue that the counterpart is an agent, and the aggressive bot's policy responds by probing less, falling to $4.9\%$ of turns at $\gamma=1.0$. Under sharp detection, the only safe behavior is the conservative one.

Probing for intelligence inherently increases the risk of detection. In the model, probe actions generate the strongest Red signal, amplifying suspicion while reducing signals associated with legitimate victims. When scammer detection capability ($\gamma$) is low, this signal cannot be reliably interpreted, allowing aggressive probing to produce higher intelligence yield. As $\gamma$ increases, however, probing becomes easier to identify, enabling scammers to screen out baiters earlier. Consequently, aggressive strategies lose their advantage, while conservative strategies trade lower initial yield for reduced detectability. At high levels of scammer detection, both approaches converge to similar outcomes.

\subsection{Mimicry Outpaces Detection in the Modeled Arms Race}
\label{sec:results_arms_race}

\begin{figure}[t]
  \centering
  \includegraphics[width=0.85\linewidth]{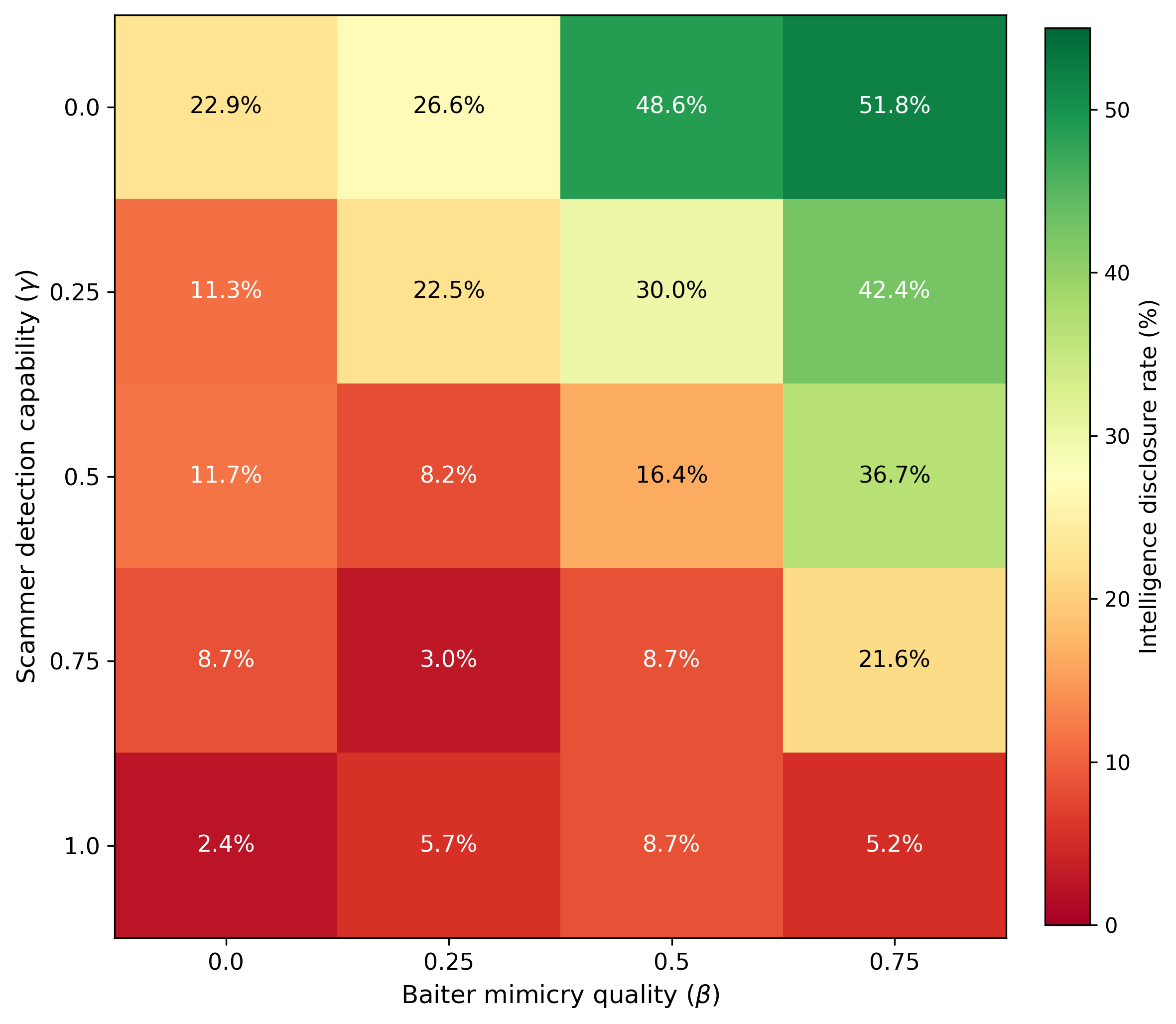}
  \caption{Intelligence disclosure rate across the
  $(\gamma, \beta)$ capability grid}
  \label{fig:rq3_arms_race2}
\end{figure}

The previous subsection varied scammer detection capability while holding baiter mimicry fixed. A more realistic threat model asks what happens when both sides improve at the same time. As automated baiting becomes more common, scammers have an incentive to search for more reliable cues that distinguish baiters from genuine victims. Defenders, in turn, have an incentive to reduce those cues by making baiter behavior more victim-like. We sweep both $\gamma$ and the baiter mimicry parameter $\beta$ from Section~\ref{sec:foundation} across the grid $[0, 0.25, 0.5, 0.75]$, extending $\gamma$ to $1.0$ to include the detection limit. For each $(\gamma, \beta)$ pair we recompute equilibrium strategies and simulate $N{=}2000$ episodes, producing the capability grid in Figure~\ref{fig:rq3_arms_race2}.

The grid's two edges show the range of outcomes under unilateral improvement, while the bottom row shows what happens at the limit of detection. In the top row, where $\gamma{=}0$ and only $\beta$ varies, the scammer's detection capability is fixed while the defender invests in mimicry, and IDR rises from $22.9\%$ at $\beta{=}0$ to $51.8\%$ at $\beta{=}0.75$, more than doubling. In the left column, where $\beta{=}0$ and only $\gamma$ varies, the opposite holds, IDR falls from $22.9\%$ at $\gamma{=}0$ to $2.4\%$ at $\gamma{=}1.0$, a reduction of nearly a factor of ten. The bottom row, at $\gamma{=}1.0$, shows that this collapse is only partial once mimicry is introduced. IDR recovers from $2.4\%$ at $\beta{=}0$ to $5.7\%$ at $\beta{=}0.25$ and $8.7\%$ at $\beta{=}0.5$, so even at the limit of scammer detection capability mimicry provides meaningful protection. A scammer who can read signals with perfect sharpness still gains little if those signals have already been made to resemble a genuine victim's.

The diagonal of the grid traces equal-pace improvement by both sides and lies between these two edges, though not at their midpoint. Starting at $22.9\%$ at $(0, 0)$, IDR rises along the diagonal to $30.0\%$ at $(\gamma{=}0.25, \beta{=}0.5)$ and to $36.7\%$ at $(\gamma{=}0.5, \beta{=}0.75)$. At the latter point, the scammer has advanced detection to $\gamma{=}0.5$, yet IDR is $36.7\%$, compared to $11.7\%$ when the scammer improves to the same level without any corresponding investment by the defender. It follows that when both sides improve at the same pace, the net effect favors the defender rather than the scammer.

The asymmetry arises because $\gamma$ and $\beta$ do not operate on the same quantity. $\beta$ alters the signal distribution the scammer observes, while $\gamma$ determines how sharply the scammer reads that distribution. When $\beta$ is high, sharper inference operates on signals that already resemble a victim's, so the marginal value of higher $\gamma$ is limited. Under this parameterization, improvements in $\beta$ erode the value of equivalent improvements in $\gamma$, which is why the diagonal tracks toward the defender-only row rather than the center of the grid.

The practical implication is that defenders must continuously improve mimicry quality to keep pace with scammer detection. Matching the scammer's capability growth increases IDR, whereas falling behind causes a sharp collapse in intelligence yield. The arms race is therefore asymmetric: the penalty for failing to improve mimicry is larger than the benefit gained from detection improvements alone.

While the IDR grid captures intelligence outcomes, Table~\ref{tab:arms_race_payoff} shows that scammer utility is comparatively insensitive to capability changes. Across all trajectories, scammer payoff varies only from $12.13$ to $20.05$ while IDR ranges from $2.4\%$ to $51.8\%$. Detection improvements alone drastically reduce defender intelligence collection but provide little additional revenue to scammers. Along the equal-pace trajectory, scammer payoff increases only modestly, as improved mimicry prolongs engagements and generates more per-turn revenue. Overall, the arms race primarily affects intelligence collection rather than scammer profitability.

\begin{table}[t]
\centering
\caption{Scammer payoff and IDR along three capability trajectories. 
Defender-only: $\gamma{=}0$, $\beta$ increases. Scammer-only: 
$\beta{=}0$, $\gamma$ increases. Equal-pace: both improve 
simultaneously. $N{=}2000$ episodes per cell.}
\label{tab:arms_race_payoff}
\scriptsize
\setlength{\tabcolsep}{4pt}
\begin{tabular}{l c c r r}
\hline
Trajectory & $\gamma$ & $\beta$ & IDR (\%) & Scammer Payoff \\
\hline
Defender only  & 0.00 & 0.00 & 22.9 & 14.81 \\
               & 0.00 & 0.25 & 26.6 & 12.13 \\
               & 0.00 & 0.50 & 48.6 & 19.97 \\
               & 0.00 & 0.75 & 51.8 & 20.05 \\
\hline
Scammer only   & 0.00 & 0.00 & 22.9 & 14.81 \\
               & 0.25 & 0.00 & 11.3 & 14.85 \\
               & 0.50 & 0.00 & 11.7 & 15.82 \\
               & 0.75 & 0.00 &  8.7 & 14.84 \\
               & 1.00 & 0.00 &  2.4 & 13.28 \\
\hline
Equal pace     & 0.00 & 0.00 & 22.9 & 14.81 \\
               & 0.10 & 0.25 & 23.3 & 13.40 \\
               & 0.25 & 0.50 & 30.0 & 15.97 \\
               & 0.50 & 0.75 & 36.7 & 18.41 \\
\hline
\end{tabular}
\end{table}

\subsection{Scaling Changes the Pace of Scambaiting, Not Whether It Works}
\label{sec:results_scaling}

Individual conversations determine intelligence yield, but campaign-level outcomes depend on how fast the scammer can process contacts and how thinly the defense is spread. We vary three scaling mechanisms independently and together. Throughput
scaling increases the number of contacts the scammer initiates per day.
Cost scaling reduces the scammer's per-turn operating cost. Coverage
scaling thins the scambaiter prevalence by spreading the same number of
agents across a larger contact pool. Each mechanism is scaled by factors
of 1, 2, 4, and 8 from a baseline campaign with 1000 contacts per day,
50 mule accounts, and 1000 email identities. The simulation runs each
campaign until a resource pool is exhausted or a time limit is reached.

Every one of the 16 runs across all four sweeps terminates the same way. The scammer exhausts its 50 mule accounts while the 1000 email identities never bind, a structural consequence of the parameterization in which mules are expensive to replace and email addresses are cheap and abundant.

Throughput scaling changes only the clock. At $1\times$ the campaign
takes 0.68 days to exhaust its mules. At $8\times$ it takes 0.08 days.
Episode-level outcomes are unchanged across all four throughput values.
IDR holds near $24\%$, abandonment near $14\%$, and mean conversation
length near 10 turns. Processing contacts faster depletes the same
resource pool faster without changing what happens in each conversation.
The other two mechanisms change episode-level outcomes in different ways, with the key metrics reported in Table~\ref{tab:rq0_episode}. The $1\times$ rows are campaign averages and differ slightly from the Table~\ref{tab:baseline_summary} baseline for this reason.

\begin{table}[t]
\centering
\caption{Episode-level outcomes under cost and coverage scaling}
\label{tab:rq0_episode}
\scriptsize
\setlength{\tabcolsep}{4pt}
\begin{tabular}{l c r r r r}
\hline
Mechanism & Scale & IDR (\%) & Abandon (\%) & Turns & Scammer payoff \\
\hline
Cost      & 1$\times$ & 21.8 & 14.0 &  9.9 & 29.3 \\
          & 2$\times$ & 24.6 &  6.1 & 11.4 & 31.3 \\
          & 4$\times$ & 36.8 &  0.0 & 12.8 & 33.7 \\
          & 8$\times$ & 32.1 &  0.0 & 13.4 & 40.1 \\
\hline
Coverage  & 1$\times$ & 16.4 & 14.5 &  9.9 & 27.3 \\
          & 2$\times$ & 30.3 &  4.5 &  9.5 & 41.0 \\
          & 4$\times$ & 36.2 &  0.7 &  9.3 & 48.1 \\
          & 8$\times$ & 39.4 &  0.2 &  9.2 & 51.8 \\
\hline
\end{tabular}
\end{table}

Cost scaling increases IDR from $21.8\%$ to $32.1\%$ at $8\times$ and eliminates abandonment beyond $4\times$. Lower per-turn costs allow scammers to remain engaged longer, increasing average conversation length from $9.9$ to $13.4$ turns and creating more opportunities for disclosure. Although individual conversations last longer, the resulting increase in IDR accelerates mule depletion and shortens campaign duration (Table~\ref{tab:rq0_campaign}). Coverage scaling increases IDR from $16.4\%$ to $39.4\%$ at $8\times$ and likewise eliminates abandonment, but through a different mechanism. As the same number of agents is distributed across a larger contact pool, effective scambaiter prevalence ($\rho$) decreases, reducing incentives for aggressive screening. This nearly doubles scammer payoff from $27.3$ to $51.8$ and allows almost four times as many episodes before mule exhaustion (Table~\ref{tab:rq0_campaign}). While the defense ultimately captures the same set of mules, the scammer extracts substantially more value before campaign termination.

\begin{table}[t]
\centering
\caption{Campaign-level outcomes under three scaling mechanisms}
\label{tab:rq0_campaign}
\scriptsize
\setlength{\tabcolsep}{4pt}
\begin{tabular}{l c r r r}
\hline
Mechanism & Scale & Episodes & Days & Scammer payoff \\
\hline
Throughput & 1$\times$ & 681 & 0.68 & 23.8 \\
           & 8$\times$ & 672 & 0.08 & 23.7 \\
\hline
Cost       & 1$\times$ & 758 & 0.76 & 29.3 \\
           & 8$\times$ & 543 & 0.54 & 40.1 \\
\hline
Coverage   & 1$\times$ & 1077 & 1.08 & 27.3 \\
           & 8$\times$ & 3934 & 3.93 & 51.8 \\
\hline
\end{tabular}

\end{table}

Throughput and cost scaling change how fast a campaign runs and how each conversation unfolds, but they do not change the fundamental constraint. Coverage scaling does. It is the only mechanism that lets the scammer extend the campaign, raise per-episode earnings, and process more victims before the mule pool runs out.

\subsection{Persistent Scammers Are the Most Productive Targets}
\label{sec:results_persistence}

Scammers differ in how long they stay in a conversation before cutting their losses. Some disengage at the first sign of an agent. Others persist for many turns even when signals are unfavorable. This subsection asks how that variation affects intelligence yield 
and scammer earnings.

We evaluate three scammer profiles. A \emph{cautious} scammer exits quickly when a conversation looks unpromising. A \emph{baseline} scammer follows the equilibrium stopping rule from Section~\ref{sec:formal_model}. A \emph{persistent} scammer stays engaged at lower beliefs about the counterpart's type, tolerating more scambaiter-like signals before abandoning. For each profile we hold the defender's policy fixed at its baseline and let the scammer recompute a best response. We test both the aggressive and conservative defender postures from Section~\ref{sec:results_detection}, with results reported in Tables~\ref{tab:rq4_aggressive} and~\ref{tab:rq4_conservative}.

\begin{table}[t]
\centering
\caption{Scammer profiles under the aggressive defender posture}
\label{tab:rq4_aggressive}
\scriptsize
\setlength{\tabcolsep}{4pt}
\begin{tabular}{l r r r r r}
\hline
Profile & IDR (\%) & Intel/ep & Abandon (\%) & Turns & Payoff \\
\hline
Cautious   &  9.9 & 0.031 & 32.9 &  8.4 & 21.95 \\
Baseline   & 20.6 & 0.058 & 13.5 & 10.0 & 29.86 \\
Persistent & 33.1 & 0.097 &  0.0 & 13.2 & 31.72 \\
\hline
\end{tabular}
\end{table}

\begin{table}[t]
\centering
\caption{Scammer profiles under the conservative defender posture}
\label{tab:rq4_conservative}
\scriptsize
\setlength{\tabcolsep}{4pt}
\begin{tabular}{l r r r r r}
\hline
Profile & IDR (\%) & Intel/ep & Abandon (\%) & Turns & Payoff \\
\hline
Cautious   &  5.8 & 0.017 & 32.1 &  8.4 & 22.91 \\
Baseline   & 14.2 & 0.041 & 14.8 &  9.8 & 29.07 \\
Persistent & 22.2 & 0.066 &  0.0 & 13.2 & 28.59 \\
\hline
\end{tabular}

\end{table}

Under both postures, persistent scammers yield the most intelligence. Under the aggressive posture, moving from cautious to persistent increases IDR from $9.9\%$ to $33.1\%$, a factor of $3.3\times$. Intel per episode triples from $0.031$ to $0.097$. Under the conservative posture, the same shift raises IDR from $5.8\%$ to $22.2\%$, a factor of $3.8\times$. A persistent scammer stays in agent conversations longer instead of abandoning them, giving the agent more turns to reach the disclosure stage.

The profitability picture is less straightforward. Persistent scammers earn slightly more than baseline under the aggressive posture ($31.72$ versus $29.86$), while under the conservative posture they earn slightly less ($28.59$ versus $29.07$). The difference comes from victim-side returns. Under the aggressive posture, victim-side payoff holds steady across profiles because the scambaiter's probing behavior does not affect victim conversations. Under the conservative posture, the persistent scammer's willingness to stay longer in all conversations reduces victim-side monetization enough to offset the gains from avoiding early abandonment.

Persistence and profitability can move in opposite directions. A scammer who never abandons generates the most intelligence for the defender without necessarily earning more. From the defender's perspective, persistent scammers are the most productive targets. From the scammer's perspective, persistence is not always the best strategy.

\section{Discussion and Conclusion}
\label{sec_conclusion}

We modeled scammer--scambaiter interactions as a dynamic Bayesian game in which scammers continuously update beliefs about the target and decide whether to continue, probe, abandon, or attempt extraction. The results show that scambaiting is fundamentally a strategic problem shaped by uncertainty, screening, resource expenditure, and information disclosure. While low deployment levels primarily generate threat intelligence, sufficiently widespread deployment can alter attacker economics by increasing the cost of false positives and wasted effort.

The analysis yields several implications for the design of scambaiting systems. First, increasing the number of baiting agents is effective only if those agents can survive early screening and reach the disclosure stage. Second, realism and low-suspicion engagement should be treated as core design objectives, since improved scammer detection reduces the effectiveness of aggressive baiting strategies. Third, defender resources should be concentrated on engagements that are likely to progress, as valuable intelligence is typically obtained only in later stages. Finally, ecosystem-level disruption requires coordinated deployment across organizations; isolated operators can collect intelligence, but broad economic impact requires scale.

The model has several limitations. Scammers are treated as independent decision makers, whereas real operations often share infrastructure, identities, and operational knowledge. The target population is also simplified to victims and baiters, omitting variation in behavior, expertise, and compliance. In addition, defender strategies remain fixed after equilibrium computation. Future work should examine networked scam operations, heterogeneous target populations, and adaptive baiting agents that learn across engagements while preserving operational safety.

More broadly, the findings should be interpreted as evidence about strategic incentives rather than predictions of any specific scam ecosystem. Empirical validation through controlled deployments will be necessary to assess how well the observed behavioral thresholds generalize to real-world environments.

\bibliographystyle{plain}
\bibliography{ecrime_2026.bib}

\appendix
\section{Model Specification}
This appendix provides the complete mathematical details underlying the
probabilistic scambaiting model summarized in Section~\ref{sec:formal_model}.
It formalizes the hidden counterparty type and belief update
(\S\ref{app:hidden_type}), the baseline signal likelihoods and capability
parameters (\S\ref{app:signal_model}), the strategy spaces for both players
(\S\ref{app:strategies}), the stage-based transition dynamics
(\S\ref{app:transition_model}), the episode utilities and terminal payoffs
(\S\ref{app:utilities}), the equilibrium characterization
(\S\ref{app:equilibrium}), and the policy computation procedure
(\S\ref{app:policy_computation}).

\begin{table}[h]
\centering
\caption{Simulation run settings and metric definitions used to generate the experimental results.}
\label{tab:sim_params}
\small
\begin{tabular}{p{0.33\linewidth} p{0.60\linewidth}}
\hline
Item & Definition / summary \\
\hline
Episodes per setting ($N$) &
Monte Carlo episodes simulated per configuration; we use $N=2000$ episodes per setting. \\

Episode horizon ($T_{\max}$) &
Maximum number of action turns per episode; default is $T_{\max}=20$. \\

Campaign length &
Maximum episodes in a campaign run; we use 5000 episodes per campaign (or earlier termination if a resource pool is exhausted). \\

Initial defender email identity pool &
Starting number of defender email identities available at campaign start; default is 1000. \\

Initial scammer money-mule payment account pool &
Starting number of scammer payment accounts available at campaign start; default is 50. \\

IDR &
Intel rate conditional on agent encounters, computed as $\#\text{intel captured} / \#\text{agent episodes}$. \\

Payment rate &
Payment extraction rate conditional on victim encounters, computed as $\#\text{payments extracted} / \#\text{victim episodes}$. \\

Abandon rate &
Fraction of episodes terminating via abandon, computed as $\#\text{abandon} / N$. \\

Mean turns &
Average number of action turns per episode, computed as the mean of $2\times|\text{turns}|$ across episodes. \\
\hline
\end{tabular}

\end{table}

\subsection{Hidden Type and Belief Update}
\label{app:hidden_type}

We model each engagement as an interaction between a scammer and a counterparty
of hidden type
\[
\theta \in \{\text{Victim},\; \text{Baiter}\}.
\]
The scammer does not observe $\theta$ directly.  It begins with a prior belief
\[
p_0 = \mathbb{P}(\theta=\text{Victim}) = 1 - \rho,
\]
where $\rho = \mathbb{P}(\theta=\text{Baiter})$ is the population-level agent
prevalence.

At each round $t$, the interaction produces a coarse conversational signal
\[
\sigma_t \in \{\text{Green},\;\text{Yellow},\;\text{Red}\},
\]
where Green denotes behavior consistent with a legitimate victim, Yellow is
ambiguous, and Red raises suspicion of defensive engagement.

\paragraph{Standard Bayesian update.}
Given prior $p_t$ and observed signal $\sigma_t$, the baseline belief update
follows Bayes' rule:
\begin{equation}
p_{t+1}
=
\frac{P(\sigma_t \mid \text{Victim})\,p_t}
     {P(\sigma_t \mid \text{Victim})\,p_t
      + P(\sigma_t \mid \text{Baiter})\,(1-p_t)}.
\label{eq:bayes_update}
\end{equation}

\paragraph{Alpha-sharpened update (scammer detection capability).}
To model improved scammer screening without changing the actual signal
distribution, we introduce a detection-sharpness parameter $\alpha > 0$.  The
likelihoods are raised to the power $\alpha$ before normalization:
\begin{equation}
p_{t+1}
=
\frac{P(\sigma_t \mid \text{Victim})^{\alpha}\,p_t}
     {P(\sigma_t \mid \text{Victim})^{\alpha}\,p_t
      + P(\sigma_t \mid \text{Baiter})^{\alpha}\,(1-p_t)}.
\label{eq:alpha_update}
\end{equation}
Setting $\alpha=1$ recovers standard Bayes' rule.  Values $\alpha>1$ make the
scammer's inference sharper (signals are treated as more informative), while
$\alpha < 1$ models a scammer that under-exploits the available evidence.  The
environment's signal-generating process is unchanged; only the scammer's
inference changes.

\subsection{Baseline Signal Likelihoods and Capability Parameters}
\label{app:signal_model}

\paragraph{Baseline likelihoods}
The baseline signal likelihoods used in the simulation experiments are:
Green: $P(\sigma \mid \text{Victim})=0.70$, $P(\sigma \mid \text{Baiter})=0.20$;
Yellow: $P(\sigma \mid \text{Victim})=0.25$, $P(\sigma \mid \text{Baiter})=0.50$;
Red: $P(\sigma \mid \text{Victim})=0.05$, $P(\sigma \mid \text{Baiter})=0.30$. These values are stylized modeling
assumptions chosen to produce weak but non-trivial signal separability, not
parameters estimated from empirical data. The specific values were selected
according to three design criteria. First, victims should be meaningfully more
likely to generate Green signals than baiters. Second, baiters should be
meaningfully more likely to generate Yellow or Red signals. Third, the gap
between the two distributions should be small enough that signals are only
weakly informative per turn, so that the scammer's belief updates gradually
over multiple turns rather than resolving type identity rapidly. If separability
were large, scammers would detect agents within one or two turns and the screening
problem would be degenerate. If separability were negligible, beliefs would
barely update and engagement dynamics would reduce to a random walk. The chosen
values place the model in the interesting intermediate regime where sustained
engagement is both necessary and uncertain. The signal model is intended to capture directional differences in observable behavior rather than to estimate exact detection probabilities. Defender actions are grouped into five categories. Within each category, the base signal probabilities are multiplied by the modifiers in Table~\ref{tab:action_modifiers} and then renormalized to sum to one. This captures how specific defender choices shift the observable signal distribution while preserving a common signal vocabulary across the game.

\begin{table}[h]
\centering
\caption{Multiplicative signal modifiers by defender action 
category, applied before renormalization. Probing actions 
strongly increase the Red signal probability; compliant and 
emotional actions increase Green.}
\label{tab:action_modifiers}
\small
\begin{tabular}{l|c|c|c}
\hline
\textbf{Action category} & \textbf{Green} & \textbf{Yellow} & \textbf{Red} \\
\hline
Emotional  & 1.3 & 0.9 & 0.7 \\
Compliant  & 1.3 & 0.9 & 0.7 \\
Stalling   & 0.8 & 1.4 & 0.9 \\
Probing    & 0.5 & 0.8 & 1.8 \\
Neutral    & 1.0 & 1.0 & 1.0 \\
\hline
\end{tabular}

\end{table}

Each baiter action in Table~\ref{tab:action_spaces} is assigned to exactly
one category: \emph{emotional} = \{\textit{show\_concern},
\textit{express\_trust}, \textit{show\_fear}\}; \emph{compliant} =
\{\textit{part\_comply}, \textit{provide\_info},
\textit{ask\_payment\_instructions}\}; \emph{stalling} =
\{\textit{tech\_confuse}, \textit{ask\_clarification},
\textit{show\_hesitation}, \textit{give\_excuse}\}; \emph{probing} =
\{\textit{request\_details}, \textit{probe\_account}\}; and
\emph{neutral} = \{\textit{process\_clarify}, \textit{express\_confusion},
\textit{apologize}\}. This categorization applies only to baiter/victim
actions; scammer actions are not assigned a signal category.

\paragraph{Beta mimicry (defender blending capability).}
The parameter $\beta \in [0,1]$ models the Baiter's ability to mimic victim
behavior.  A higher $\beta$ shifts the Baiter's signal distribution toward the
Victim distribution, making the two types harder to distinguish.  Formally, for
each signal $\sigma$:
\begin{equation}
P_{\beta}(\sigma \mid \text{Baiter})
= (1-\beta)\,P_0(\sigma \mid \text{Baiter})
  + \beta\,P_0(\sigma \mid \text{Victim}),
\end{equation}
where $P_0(\cdot\mid\cdot)$ denotes the baseline likelihoods defined above.

\paragraph{Gamma separability (information structure quality).}
The parameter $\gamma \in [0,1]$ captures how type-revealing the conversational
channel is, independently of either player's strategy.  Higher $\gamma$
increases the statistical separation between type-conditional signal
distributions for \emph{both} types:
\begin{align}
P_{\gamma}(\sigma \mid \text{Victim})
&= (1-\gamma)\,P_0(\sigma \mid \text{Victim})
   + \gamma\,\mathbf{1}\{\sigma=\text{Green}\}, \\
P_{\gamma}(\sigma \mid \text{Baiter})
&= (1-\gamma)\,P_0(\sigma \mid \text{Baiter})
   + \gamma\,\mathbf{1}\{\sigma=\text{Red}\}.
\end{align}
At $\gamma=0$ the baseline distributions are preserved; at $\gamma=1$ each type
emits a perfectly diagnostic signal.  Unlike $\alpha$, which changes only the
scammer's inference, $\gamma$ changes the underlying information structure.

\subsection{Player Strategies}
\label{app:strategies}

Let $A_{\mathrm{SB}}(s)$ and $A_{\mathrm{DB}}(s)$ denote the scammer and
defender action sets in state $s$ (Table~\ref{tab:action_spaces}).

\begin{table*}[t]
\centering
\caption{Action spaces for both sides by state.}
\label{tab:action_spaces}
\small
\begin{tabular}{p{2.2cm}|p{6.2cm}|p{6.2cm}}
\hline
\textbf{State} & \textbf{Scammer Actions} & \textbf{Baiter Actions} \\
\hline
Trust Building & establish\_problem, show\_expertise, 
abandon\_target & show\_concern, process\_clarify, 
express\_trust \\
\hline
Hook Established & allege\_fraud, create\_urgency, 
abandon\_target & show\_concern, process\_clarify, 
express\_trust \\
\hline
Request Phase & request\_payment, apply\_pressure, 
specify\_payment\_method, abandon\_target & tech\_confuse, 
ask\_clarification, give\_excuse, part\_comply, 
ask\_payment\_instructions \\
\hline
Disclosure & verify\_information, proceed\_transaction, 
share\_mule\_details, abandon\_target & provide\_info, 
show\_hesitation, request\_details, probe\_account \\
\hline
Breakdown & legal\_threat, abandon\_target & show\_fear, 
express\_confusion, apologize \\
\hline
\end{tabular}
\end{table*}

\paragraph{Victim policy.}
The Victim is modeled as non-strategic.  It follows a fixed behavioral policy
\[
\pi_{\mathrm{RV}} : S \;\to\; \Delta\!\bigl(A_{\mathrm{DB}}(s)\bigr),
\]
selecting actions in each state without optimizing over expected payoffs.
Victims do not have access to the agent-only intel actions
$\{\textit{request\_details},\;\textit{probe\_account}\}$.

\paragraph{Baiter policy.}
The Baiter is strategic and chooses actions to maximize expected defender
utility (engagement duration and intelligence capture).  Its information set is
the current kill-chain state $s_t$ and the scammer's most recent action
$a_{SB,t}$; it does \emph{not} observe the scammer's belief $p_t$.  Its policy
is therefore:
\begin{equation}
\pi_{\mathrm{B}} : S \times A_{\mathrm{SB}} \;\to\; \Delta\!\bigl(A_{\mathrm{DB}}(s)\bigr).
\end{equation}
Conditioning on the scammer's action allows the Baiter to respond to observed
escalation or probing behaviors, while the absence of direct belief observation
reflects the realistic constraint that the defender does not have access to the
scammer's internal state.

\paragraph{Scammer policy.}
The scammer's decision depends on both the current kill-chain state and its
current belief, since the belief determines whether continued investment in the
interaction is worthwhile:
\begin{equation}
\pi_{\mathrm{SB}} : S \times [0,1] \;\to\; \Delta\!\bigl(A_{\mathrm{SB}}(s)\bigr).
\end{equation}
Because the belief space is continuous, we approximate it with a discrete grid
$\mathcal{B} = \{0.05, 0.15, 0.25, 0.35, 0.45, 0.55, 0.65, 0.75, 0.85,
0.95\}$, and restrict the scammer's policy to $S \times \mathcal{B}$.

\subsection{Stage-Based Transition Dynamics}
\label{app:transition_model}

Let $S$ denote the kill-chain states $\{\text{Trust},\;\text{Hook},\;\text{Request},$ $\text{Disclosure},\;\text{Breakdown},\;\text{Termination}\}$.
The general transition kernel is
\begin{equation}
\mathcal{P}(s_{t+1} \mid s_t, p_t, a_{SB,t}, a_{DB,t}).
\label{eq:transition_kernel}
\end{equation}
All transitions use the logistic (sigmoid) function
$\sigma(z) = 1/(1+e^{-z})$.
We now specify each stage in full.

\paragraph{Abandonment (all states).}
If the scammer plays \textit{abandon\_target} in any non-terminal state, the
episode ends immediately:
\begin{align}
&\mathcal{P}(\text{Termination} \mid s_t \neq \text{Termination},\; \nonumber\\
&\qquad a_{SB,t}=\textit{abandon\_target}) = 1.
\end{align}

\paragraph{Trust Building $\to$ Hook Established.}
From Trust Building, the scammer advances with probability
\begin{equation}
\mathcal{P}(\text{Hook} \mid \text{Trust},\,p_t)
= \sigma\!\bigl(k_{\mathrm{trust}}\,(p_t - \tau_{\mathrm{trust}})\bigr),
\end{equation}
and otherwise remains in Trust Building.  The baseline parameters are
$k_{\mathrm{trust}}=5.0$ and $\tau_{\mathrm{trust}}=0.40$.  Higher belief
increases the probability of advancing because the scammer has more incentive
to invest further when it is confident about the target's type.

\paragraph{Hook Established $\to$ Request Phase.}
\begin{equation}
\mathcal{P}(\text{Request} \mid \text{Hook},\,p_t)
= \sigma\!\bigl(k_{\mathrm{hook}}\,(p_t - \tau_{\mathrm{hook}})\bigr),
\end{equation}
with $k_{\mathrm{hook}}=5.0$ and $\tau_{\mathrm{hook}}=0.40$.

\paragraph{Request Phase $\to$ Disclosure.}
In the Request Phase, the probability of advancement depends on both belief and
the defender's action category.  A compliant defender action (e.g.,
\textit{part\_comply}) signals willingness to proceed and increases the
transition probability; a stalling action (e.g., \textit{tech\_confuse})
suppresses it:

\begin{align}
&\mathcal{P}(\text{Disclosure} \mid \text{Request},\,p_t,\,a_{DB,t}) \nonumber\\
&\quad = \min\!\Bigl(1,\;
  \sigma\!\bigl(k_{\mathrm{req}}\,(p_t - \tau_{\mathrm{req}})\bigr)
  \cdot m(a_{DB,t})\Bigr),
\end{align}

where $k_{\mathrm{req}}=5.0$, $\tau_{\mathrm{req}}=0.35$, and the action
modifier $m(a_{DB,t})$ takes value $1.5$ for compliant actions, $0.5$ for
stalling actions, and $1.0$ otherwise.

\paragraph{Disclosure dynamics.}
Disclosure is a pivot state from which three outcomes compete: the episode may
remain in Disclosure, transition to Breakdown, or terminate.

Breakdown is triggered by low belief (the scammer suspects an agent) via an
\emph{inverted} sigmoid:
\begin{equation}
\mathcal{P}(\text{Breakdown} \mid \text{Disclosure},\,p_t)
= \frac{1}{1+e^{\,k_{\mathrm{disc}}\,(p_t-\tau_{\mathrm{disc}})}},
\end{equation}
with $k_{\mathrm{disc}}=5.0$ and $\tau_{\mathrm{disc}}=0.30$.  When the
scammer plays a commit action
$a_{SB,t} \in A_{\mathrm{SB}}^{\mathrm{commit}}
= \{\textit{proceed\_transaction},\;\textit{share\_mule\_details},\;
\textit{legal\_threat}\}$,
termination probability also rises with belief:

\begin{align}
&\mathcal{P}(\text{Termination} \mid \text{Disclosure},\,p_t,\,
a_{SB,t} \in A_{\mathrm{SB}}^{\mathrm{commit}}) \nonumber\\
&\quad = \sigma\!\bigl(k_{\mathrm{disc}}\,(p_t-\tau_{\mathrm{disc}})\bigr).
\end{align}

Note that $p_{\mathrm{term}} = \sigma(k_{\mathrm{disc}}(p_t - 
\tau_{\mathrm{disc}}))$ and $p_{\mathrm{bd}} = 1/(1 + 
e^{\,k_{\mathrm{disc}}(p_t - \tau_{\mathrm{disc}})})$ are 
complements: $p_{\mathrm{term}} + p_{\mathrm{bd}} = 1$ for 
all $p_t$. Combining these, when the scammer plays a commit 
action the three outcomes have probabilities:
\begin{align}
\mathcal{P}(\text{Termination} \mid \cdot) &=
  p_{\mathrm{term}}\,(1-p_{\mathrm{bd}}),\\
\mathcal{P}(\text{Breakdown} \mid \cdot)   &= p_{\mathrm{bd}},\\
\mathcal{P}(\text{Disclosure} \mid \cdot)  &=
  (1-p_{\mathrm{term}})\,(1-p_{\mathrm{bd}}),
\end{align}
which sum to one by construction. When the scammer does not 
play a commit action, the termination channel is closed and 
only the Breakdown/stay competition applies:
\begin{align}
\mathcal{P}(\text{Breakdown} \mid \cdot)  &= p_{\mathrm{bd}},\\
\mathcal{P}(\text{Disclosure} \mid \cdot) &= 1 - p_{\mathrm{bd}}.
\end{align}

\paragraph{Breakdown $\to$ Termination.}
Once the interaction enters Breakdown, it terminates in the next step with
probability one:
\begin{equation}
\mathcal{P}(\text{Termination} \mid \text{Breakdown}) = 1.
\end{equation}

\subsection{Episode Utilities and Terminal Payoffs}
\label{app:utilities}
This appendix uses expanded notation for the payoff 
parameters introduced in Table~\ref{tab:payoff_main}. 
The correspondence is $r_{\mathrm{DB}} \equiv r_{\text{occ}}$, 
$r_{\mathrm{SB}} \equiv c_b$ and $c_v$, 
$R_{\mathrm{SB}}^{\mathrm{abandon}} \equiv p_{\text{ab}}$, 
$R_{\mathrm{DB}}^{\mathrm{intel}} \equiv r_{\text{intel}}$, 
and $R_{\mathrm{SB}}^{\mathrm{pay}} \equiv r_{\text{pay}}$.

\paragraph{Per-turn stage rewards.}
For each non-terminal turn $t$, the Baiter receives a fixed reward for keeping
the scammer engaged:
\begin{equation}
r_{\mathrm{DB}} = +3.
\end{equation}
For the scammer, the per-turn cost depends on the true (unobserved) defender
type:
\begin{equation}
r_{\mathrm{SB}}(\theta) =
\begin{cases}
-5, & \theta=\text{Baiter},\\
\phantom{-}0, & \theta=\text{Victim}.
\end{cases}
\end{equation}
Because the scammer does not observe $\theta$, its expected per-turn payoff is
$r_{\mathrm{SB}}^{\mathrm{exp}} = p_t \cdot c_v + (1-p_t)\cdot c_b = c_b(1-p_t)$.

\paragraph{Terminal rewards and penalties.}
\emph{Abandonment.}  If the scammer plays \textit{abandon\_target}, the episode
ends immediately with scammer penalty $R_{\mathrm{SB}}^{\mathrm{abandon}} = -20$
and no terminal payoff for the defender.

\emph{Intelligence capture (Baiter episodes).}
Let $A_{\mathrm{DB}}^{\mathrm{intel}} = \{\textit{request\_details},\;
\textit{probe\_account}\}$.  When $s_t \in \{\text{Disclosure},\text{Breakdown}\}$,
$a_{SB,t} \in A_{\mathrm{SB}}^{\mathrm{commit}}$, and
$\theta = \text{Baiter}$, the probability of capturing actionable intelligence is:
\begin{equation}
p_{\mathrm{intel}}(s_t, a_{DB,t}) =
\begin{cases}
0.85, & s_t=\text{Disclosure},\; a_{DB,t}\in A_{\mathrm{DB}}^{\mathrm{intel}},\\
0.30, & s_t=\text{Disclosure},\; a_{DB,t}\notin A_{\mathrm{DB}}^{\mathrm{intel}},\\
0.20, & s_t=\text{Breakdown}.
\end{cases}
\end{equation}
Successful capture yields $R_{\mathrm{DB}}^{\mathrm{intel}} = +150$; the scammer
receives zero.

\emph{Payment extraction (Victim episodes).}
Let $A_{\mathrm{DB}}^{\mathrm{pay}} = \{\textit{provide\_info}\}$.  When
$s_t \in \{\text{Disclosure},\text{Breakdown}\}$,
$a_{SB,t} \in A_{\mathrm{SB}}^{\mathrm{commit}}$, and $\theta=\text{Victim}$,
the payment success probability is:
\begin{equation}
p_{\mathrm{pay}}(s_t, a_{DB,t}) =
\begin{cases}
0.70, & s_t=\text{Disclosure},\; a_{DB,t}\in A_{\mathrm{DB}}^{\mathrm{pay}},\\
0.20, & s_t=\text{Disclosure},\; a_{DB,t}\notin A_{\mathrm{DB}}^{\mathrm{pay}},\\
0.30, & s_t=\text{Breakdown}.
\end{cases}
\end{equation}
Successful extraction yields $R_{\mathrm{SB}}^{\mathrm{pay}} = +100$; the
defender receives zero on victim episodes.

\paragraph{Episode-level utilities.}
Let $T$ denote the stopping time and $\gamma_d = 0.95$ the 
discount factor. Throughout this appendix $\gamma_d$ denotes 
the discount factor exclusively; it is distinct from the 
signal separability parameter $\gamma$ defined in 
\S\ref{app:signal_model}.

The episode-level utilities are
\begin{align}
U_{\mathrm{DB}}
&= \mathbb{E}\!\left[
    \sum_{t=0}^{T-1} \gamma_d^t\,r_{\mathrm{DB}}
    \;+\; \gamma_d^T\,\mathbf{1}\{\textit{intel}\}\,R_{\mathrm{DB}}^{\mathrm{intel}}
   \right],
\label{eq:defender_utility} \\
U_{\mathrm{SB}}
&= \mathbb{E}\!\left[
    \sum_{t=0}^{T-1} \gamma_d^t\,r_{\mathrm{SB}}(\theta)
    \;+\; \gamma_d^T\!\left(
      \mathbf{1}\{\textit{pay}\}\,R_{\mathrm{SB}}^{\mathrm{pay}}
    \right.\right. \notag \\
&\qquad\qquad\qquad\left.\left.
      + \mathbf{1}\{\textit{abandon}\}\,R_{\mathrm{SB}}^{\mathrm{abandon}}
    \right)
   \right],
\label{eq:scammer_utility}
\end{align}
where expectations are over the hidden type draw (governed by $\rho$), the
signal process, state transitions, stochastic terminal outcomes, and any
strategy randomization.

\subsection{Equilibrium Characterization}
\label{app:equilibrium}

The episode game is a finite-horizon Dynamic Bayesian Game.  With $\pi_{\mathrm{RV}}$
held fixed, the strategic interaction is between the scammer and the Baiter.

The scammer's best response to a Baiter policy $\pi_{\mathrm{B}}$ is:
\begin{equation}
\pi^*_{\mathrm{SB}}
\;\in\; \arg\max_{\pi_{\mathrm{SB}}}\;
U_{\mathrm{SB}}(\pi_{\mathrm{SB}},\,\pi_{\mathrm{B}};\,\pi_{\mathrm{RV}}).
\end{equation}

The Baiter's best response to a scammer policy $\pi_{\mathrm{SB}}$ is:
\begin{equation}
\pi^*_{\mathrm{B}}
\;\in\; \arg\max_{\pi_{\mathrm{B}}}\;
U_{\mathrm{DB}}(\pi_{\mathrm{B}},\,\pi_{\mathrm{SB}};\,\pi_{\mathrm{RV}}).
\end{equation}

An \emph{approximate Nash equilibrium} for the episode game is a pair
$(\pi^*_{\mathrm{SB}},\,\pi^*_{\mathrm{B}})$ such that each policy is a best
response to the other on the discretized belief grid $\mathcal{B}$.  Because the Baiter does not observe $p_t$, it cannot 
condition its action directly on the scammer's current 
belief. Instead, for each observable scammer action 
$a_{\mathrm{SB}} \in A_{\mathrm{SB}}(s)$, the Baiter 
evaluates each of its own actions $a_{\mathrm{DB}}$ by 
taking an expectation over the belief grid $\mathcal{B}$, 
weighted by the probability that the scammer would play 
$a_{\mathrm{SB}}$ at each belief level under the current 
scammer strategy:
\begin{equation}
\begin{split}
&Q_{\mathrm{B}}(s,\, a_{\mathrm{SB}},\, a_{\mathrm{DB}},\, h)
\\&=
\frac{
  \sum_{b \in \mathcal{B}}\,
  \pi_{\mathrm{SB}}(a_{\mathrm{SB}} \mid s, b)\,
  V_{\mathrm{B}}(s,\, b,\, a_{\mathrm{SB}},\, a_{\mathrm{DB}},\, h)
}{
  \sum_{b \in \mathcal{B}}\,
  \pi_{\mathrm{SB}}(a_{\mathrm{SB}} \mid s, b)
},
\label{eq:baiter_q}
\end{split}
\end{equation}
where $V_{\mathrm{B}}$ is the Baiter's expected value of 
taking action $a_{\mathrm{DB}}$ at belief $b$ with $h$ turns 
remaining, and $h$ indexes the backward induction horizon. 
The Baiter then selects the action that maximizes 
$Q_{\mathrm{B}}$ for each $(s, a_{\mathrm{SB}})$ pair.

\subsection{Policy Computation}
\label{app:policy_computation}

\begin{algorithm}[t]
\caption{Solve-then-Simulate Pipeline}
\label{alg:pipeline}
\textbf{Require:} Strategies $(\pi_{\mathrm{SB}},\, \pi_{\mathrm{B}})$
from Section~\ref{subsec:computing_strategies};
episodes $N = 2000$; horizon $H = 25$
\begin{algorithmic}[1]
\State $\mathcal{R} \leftarrow \emptyset$
\For{episode $i = 1, 2, \ldots, N$}
    \State Sample defender type $\theta \sim \mathrm{Bernoulli}(\rho)$;
           initialize $s \leftarrow \textsc{Trust}$,\;
           $\hat{p} \leftarrow \textsc{Discretize}(1{-}\rho,\, \mathcal{B})$
    \For{turn $t = 1, 2, \ldots, H$}
        \State Sample scammer action $a_{\mathrm{SB}} \sim \pi_{\mathrm{SB}}(s,\, \hat{p})$
        \If{$a_{\mathrm{SB}} = \textit{abandon}$}
            \State Record abandonment payoffs;\quad \textbf{break}
        \EndIf
        \State Sample realized defender action $a_{\mathrm{D}}$
               from $\pi_{\mathrm{B}}(s,\, a_{\mathrm{SB}})$ if $\theta = \mathrm{Baiter}$,
               else from $\pi_{\mathrm{RV}}(s)$
        \State Generate signal $\sigma \sim P(\cdot \mid \theta,\, a_{\mathrm{D}})$
               from true type and realized action
        \State Update belief using policy-implied actions
               (scammer observes $\sigma$, not $a_{\mathrm{D}}$):
               $\hat{p} \leftarrow \textsc{Discretize}(
               \textsc{UpdateBelief}(\hat{p},\, \sigma,\, \pi_{\mathrm{B}},\,
               \pi_{\mathrm{RV}}),\, \mathcal{B})$
        \State Sample next state
               $s \sim \mathcal{P}(\cdot \mid s,\, a_{\mathrm{SB}},\,
               a_{\mathrm{D}},\, \hat{p})$
        \If{$s = \textsc{Termination}$}
            \State Sample terminal outcome stochastically;\quad \textbf{break}
        \EndIf
    \EndFor
    \State Append episode payoffs and outcomes to $\mathcal{R}$
\EndFor
\State \Return sample averages over $\mathcal{R}$
\end{algorithmic}
\end{algorithm}

Both best-response problems are solved via finite-horizon backward
induction, written $\textsc{BackwardInduction}_{\mathrm{SB}}$ and
$\textsc{BackwardInduction}_{\mathrm{B}}$ respectively, over the product
space $S \times \mathcal{B}$ for the scammer and $S \times A_{\mathrm{SB}}$
for the Baiter, with a horizon of $H = \lfloor T_{\max}/2 \rfloor = 25$
interaction steps.

\paragraph{Iterative best-response with damping}
We compute the approximate equilibrium by alternating best-response
computations until convergence. To avoid oscillation, a known failure mode
of pure iterative best-response in games with cyclic incentives, we apply a
damped update at each iteration with damping factor $\eta = 0.2$. Both
best responses are computed simultaneously before either strategy is
updated, preventing the sequential update bias that causes cycling.
Convergence is declared when the total $\ell_1$ strategy distance falls
below $\varepsilon = 10^{-4}$. We additionally detect 2-cycles: if the
strategy pair at iteration $k$ is within $\varepsilon_{\mathrm{cyc}} =
10^{-6}$ of the pair at iteration $k-2$, we resolve by averaging the last
two profiles and halting. Algorithm~\ref{alg:equilibrium} states the full
procedure; it implements the \textsc{ComputeEquilibrium} call in
Algorithm~\ref{alg:pipeline} line~1. In practice it converges in under 50
iterations for all parameter configurations reported in this paper.

\paragraph{Outcome estimation.}
The converged equilibrium strategies $(\pi_{\mathrm{SB}}^*, \pi_{\mathrm{B}}^*)$
are then used to drive Monte Carlo simulation over $N = 2000$ episodes per
experimental condition to estimate the outcome statistics reported in
Section~\ref{sec:results}.



\end{document}